\documentclass[aps,pra,reprint,amsmath,amssymb,groupedaddress,showpacs]{revtex4-1}
\usepackage{graphicx}
\usepackage{float}
\usepackage{dcolumn} 
\usepackage{bm}      
\usepackage[bookmarks=false]{hyperref} 
\usepackage[font={small}]{caption} 
\hypersetup{
    unicode=false,          
    pdftoolbar=true,        
    pdfmenubar=true,        
    pdffitwindow=true,      
    pdfstartview={},        
    pdftitle={Evidence that All States Are Unitarily Equivalent to X States of Same Entanglement}, 
    pdfauthor={Samuel R. Hedemann},
    pdfsubject={Generalizations of X States and Quantum Entanglement}, 
    pdfcreator={PDFLaTeX},  
    pdfproducer={PDFLaTeX}, 
    pdfkeywords={X States,} {True-Generalized X States,} {TGX States,} {Maximally Entangled Mixed States,} {MEMS,} {2x2,} {2x3,} {2x3 MEMS,} {Concurrence,} {X Concurrence,} {Purity,} {Concurrence-Purity,} {CP,} {Unitary,} {Local-Unitary,} {Entanglement-Preserving Unitary,} {EPU,} {Entanglement,} {Entanglement Measure,} {Quantum Channels,} {H States,} {Partial Trace,} {Partial Transpose,} {Negativity,} {Multipartite,} {N-Partite}, 
    pdfnewwindow=true,      
    colorlinks=true,        
    linkcolor=black,        
    citecolor=black,        
    filecolor=black,        
    urlcolor=black          
}
\begin{document}
\title{Evidence~that~All~States~Are~Unitarily~Equivalent~to~X~States~of~the~Same~Entanglement}
\author{Samuel R.~Hedemann}
\affiliation{Dept.~of Physics and Engineering Physics, Stevens Institute of Technology, Hoboken, NJ 07030, USA}
\date{\today}
\begin{abstract}
Strong numerical evidence is presented suggesting that all two-qubit mixed states are equivalent to X states by a single entanglement-preserving unitary (EPU) transformation, so that the concurrence of such an X state equals that of the original general state.  An X-state parameterization of a general two-qubit state is given, allowing all states to have their concurrence parametrically specified.  A new kind of entanglement measure is proposed, relating a general state's entanglement to that of a pure state in the same system.  New states called ``H States'' are presented, having fully parametric concurrence and purity, with the intention of using them to construct entanglement-preserving depolarization channels, which may aid development of the new entanglement measure.  A theory of ``true-generalized'' X states (TGX states) is proposed for the general case of $N$-partite systems.  While such states do not generally have the literal ``X'' shape, evidence is shown that they are the true generalizations of X states in larger systems, since they appear to always be EPU-equivalent to general states of all ranks, whereas literal X states generally are not.  An example of this is given for $2\times 3$, including the proposition of the $2\times 3$ maximally entangled mixed states (MEMS).  If the claim that TGX states are universal is valid, then any entanglement measure may be computable in a simpler form by using the EPU-equivalence between general states and TGX states.
\end{abstract}
\maketitle
\section{\label{sec:I}Introduction}
In the field of quantum information, much work has been done showing the benefits of preparing systems in states with special simple forms, such as X states (to be defined below), in which many of the density-matrix elements are zero.  Experimentally, this simplicity can reduce equipment, complexity, time, and cost.  From a theoretical viewpoint, such simple states can allow \textit{symbolic} computation of entanglement, an advantage because entanglement of general states must usually be computed numerically, which can obscure the algebraic dependence of entanglement upon various parameters of interest.  Therefore, it would be highly beneficial, both experimentally and theoretically, if we could somehow convert any given general state to a simple X state of the same entanglement.

The most important result of this paper is that it presents \textit{strong numerical evidence showing that any general state can indeed be transformed to an X state of the same entanglement}.  Section~\ref{sec:II} of this paper provides this numerical evidence for two-qubit systems, the simplest case. Then, Sec.~\ref{sec:III} generalizes the idea to $N$-partite systems, giving numerical evidence for $2\times 3$ systems as an example.  Since an explicit general multipartite entanglement measure has not yet been discovered beyond $2\times 3$ systems, Sec.~\ref{sec:II} also suggests a new kind of entanglement measure that may enable further progress in the multipartite case, introducing a new two-qubit parameteric state family called ``H states'' which may be useful for this task if they can be generalized.  Some important concepts for this paper are concurrence, X states, and maximally entangled mixed states (MEMS), all of which we now review before proceeding.

Hill and Wootters' landmark papers \cite[]{HiWo,Woot} presented the \textit{concurrence}, a tool that allows the quantification of entanglement in any mixed two-qubit quantum state defined by the density matrix $\rho =\sum_{j}p_{j}\rho_{j}$, where $p_{j}$ are probabilities such that $\sum_{j}p_{j} =1$ and $p_{j}\in[0,1]$, and the $\rho_{j}$ are pure states.  Originally developed to calculate the entanglement of formation \cite[]{EntF}, the concurrence is a measure of entanglement as well, and is defined as
\begin{equation}
C(\rho ) \equiv \max \{0,\lambda _1  - \lambda _2  - \lambda _3  - \lambda _4 \},
\label{eq:1}
\end{equation}
where $\{ \lambda_{1},\ldots,\lambda_{4}\}$ are the eigenvalues, in decreasing order, of the Hermitian matrix $R \equiv \sqrt {\sqrt \rho  \tilde \rho \sqrt \rho  } $, where 
\begin{equation}
\tilde{\rho}  \equiv (\sigma_2  \otimes \sigma _2 )\rho ^* (\sigma _2  \otimes \sigma _2 ),
\label{eq:2}
\end{equation}
where $\sigma_2$ is a Pauli matrix.  The $\lambda_{k}$ are also the square-roots of the eigenvalues of the non-Hermitian matrix $\rho \tilde{\rho}$.  Due to the reliance on eigenvalues of transformed quantities, the concurrence is generally only suited to numerical calculations, though it is still highly useful.

However, a useful special case was presented in Yu and Eberly's work on the concurrence of X states \cite[]{YuEb}, defined as those states with the form
\begin{equation}
\rho  = \left( {\begin{array}{*{20}c}
   {\rho _{1,1} } & 0 & 0 & {\rho _{1,4} }  \\
   0 & {\rho _{2,2} } & {\rho _{2,3} } & 0  \\
   0 & {\rho _{3,2} } & {\rho _{3,3} } & 0  \\
   {\rho _{4,1} } & 0 & 0 & {\rho _{4,4} }  \\
\end{array}} \right)\!,
\label{eq:3}
\end{equation}
for which the concurrence was shown to be
\begin{equation}
C(\rho ) = 2\max \{ 0,|\rho _{3,2} | - \sqrt {\rho _{4,4} \rho _{1,1} } ,|\rho _{4,1} | - \sqrt {\rho _{3,3} \rho _{2,2} } \},
\label{eq:4}
\end{equation}
which holds true for both pure and \textit{mixed} X states.  Since many well-known and useful families of states have X form, including the Bell states, Werner states \cite[]{Wern}, and isotropic states, (\ref{eq:4}) is a powerful tool for obtaining symbolic expressions for the concurrence of X states.

Furthermore, it was shown in \cite[]{Rafs} that the X-state part of any state, along with the anti-X matrix formed by looking at the remaining elements outside of the X form, can be used to find lower bounds on concurrence for non-X states.  The same work also shows that literal X states in larger systems can be used to find lower bounds on generalizations of concurrence such as $I$ concurrence \cite[]{Rung}, though this is limited since such measures are merely \textit{sufficient} to detect entanglement.

Much work has also been done in exploring the relationship of concurrence as a function of purity $P(\rho)\equiv\text{tr}(\rho^{2})$ such as in \cite[]{Ishi,Hors,Zima}, which discuss the idea of maximally entangled mixed states (MEMS), the latter of which investigates the action of local channels on X states.  MEMS are states with the maximum entanglement possible for a given purity, and are defined in \cite[]{Zima} for two qubits as all states local-unitarily equivalent to
\begin{equation}
\rho _{\text{MEMS}}  \equiv\! \left\{\! {\begin{array}{*{20}l}
   \!\!{\left(\!\! \begin{array}{l}
 p|\Phi ^ +  \rangle \langle \Phi ^ +  | + {\textstyle{1 \over 3}}|0,1\rangle \langle 0,1| \\ 
  + ({\textstyle{1 \over 3}} - {\textstyle{p \over 2}})\left(\! \begin{array}{l}
 |0,0\rangle \langle 0,0| \\ 
  + |1,1\rangle \langle 1,1| \\ 
 \end{array}\! \right) \\ 
 \end{array}\!\! \right)\!\!;} & {p \in [0,{\textstyle{2 \over 3}}]}  \\
   {p|\Phi ^ +  \rangle \langle \Phi ^ +  | + (1 - p)|0,1\rangle \langle 0,1|;} & {p \in [{\textstyle{2 \over 3}},1],}  \\
\end{array}} \right.
\label{eq:5}
\end{equation}
where $|\Phi ^ +  \rangle  \equiv {\textstyle{1 \over {\sqrt 2 }}}(|0,0\rangle  + |1,1\rangle )$ is a Bell state.  The concurrence-purity (CP) plot of $\rho _{\text{MEMS}}$ gives the maximum entanglement possible for \textit{all} mixed two-qubit states.  Note that the $\rho _{\text{MEMS}}$ shown in (\ref{eq:5}) are X states.

In this paper, we will see strong evidence that for two qubits, X states can access all possible concurrence and purity (CP) combinations, and that for every general state $\rho_G$, there is an X state to which it can be transformed using a single entanglement-preserving unitary (EPU) matrix.  Such EPU transformations are not necessarily a tensor product of unitary matrices, and a more general form for them will be given in Sec.~\ref{sec:II} and App.~\ref{sec:AppA}.

It will then be shown that if the above transformation is always possible, then general density matrices can be parameterized entirely in terms of an X state transformed by a single EPU matrix, so that Yu and Eberly's formula in (\ref{eq:4}) allows us to parametrically specify the concurrence of any general mixed state through its X-state core.

Then, a new kind of entanglement measure will be briefly proposed, based on relating the concurrence of any state to that of a pure state in the same system.  To assist in the development of this entanglement measure, this paper will also present a new type of state that has fully parametric concurrence and purity.

Finally, to generalize these ideas, in Sec.~\ref{sec:III} we will define and examine methods to find ``true-generalized X states'' (TGX states), for all discrete quantum systems, which yield TGX states even in systems for which literal X states are not the same as TGX states, as in $2\times 3$.  The hypothesis that TGX states are universally equivalent to all quantum states up to an EPU transformation is further supported by examples in $2\times 3$, and is conjectured to hold true for all multipartite systems.
\section{\label{sec:II}Universality of X States for Two-Qubit Systems}
Here, we focus only on two-qubit systems (having $n=4$ dimensions), with the goals of showing that all states are EPU-equivalent to X states and of finding an X-state parameterization for all two-qubit states.  A new kind of entanglement measure is also proposed, and new states called H states are presented, having the feature of parametric concurrence and purity.

For most physicists, the new term ``entanglement-preserving unitary'' (EPU) probably calls to mind the local-unitary operations, which are tensor products of unitary operators in each subsystem, such as $U^{(1)}\otimes U^{(2)}$, where $U^{(m)}$ is a unitary operator in subsystem $m$.  Indeed, local-unitary matrices \textit{do} qualify as EPU.  However, the special form of X states actually enables a \textit{wider} class of unitary matrices to be entanglement-preserving, as is proved in App.~\ref{sec:AppA}.  Therefore, local-unitarity is merely \textit{sufficient} for ensuring entanglement preservation, and thus we need a more general term for more general cases.  The most transparent term for these is ``entanglement-preserving unitary'' (EPU) transformations.  In general, labeling these EPU transformations as $U_{\text{EPU}}$, they can be formally defined, for a given input state $\rho$, by
\begin{equation}
U_{\text{EPU}}  \equiv U,\;\;\text{s.t.}\;\;\left\{\! {\begin{array}{*{20}l}
   {U^{\dag}  } &\!\! { = U^{ - 1} }  \\
   {E(U\rho U^{\dag}  )} &\!\! { = E(\rho ), }  \\
\end{array}} \right.
\label{eq:6}
\end{equation}
where $E(\rho)$ is any valid entanglement measure and $\rho$ is any state.  Thus, (\ref{eq:6}) means that $U_{\text{EPU}}$ is unitary, and that given input state $\rho$, the transformed state $U_{\text{EPU}}\rho U_{\text{EPU}}^{\dag}$ has the same entanglement as $\rho$.

First, note that the particular entanglement measure used in (\ref{eq:6}) is not important because all that matters is that its value is unchanged by the application of $U_{\text{EPU}}$.  Secondly, notice that if the set of states $\rho$ is restricted to a particular type of state, such as the X states, then that can affect the definition of \textit{which} unitary matrices qualify as EPU, due to the dependence of (\ref{eq:6}) on $\rho$.

As proven in App.~\ref{sec:AppA}, the set of EPU matrices acting only on X states includes \textit{nonlocal} unitary matrices, so that these EPU matrices will generally \textit{not} have product-form.  See Sec.~\ref{sec:II.D.1} for a compact summary of the form of these EPU matrices, or see App.~\ref{sec:AppA} for more details.

As a preview of the most important results of this section, the main numerical evidence that shows that there always exists an EPU matrix that transforms a general two-qubit mixed state into an X state of the same concurrence is given in Fig.~\ref{fig:2} compared to Fig.~\ref{fig:1}.  These show that it is always possible to find an X state with the same concurrence-purity-rank (CPR) combination as any general state.  Then, given that numerically-supported fact, (\ref{eq:17}) shows how to find the EPU matrix that causes the desired transformation, while Fig.~\ref{fig:3} shows that it works on a large number of consecutive arbitrary mixed states.
\subsection{\label{sec:II.A}X States Contain Maximal Concurrence-Purity Combinations for All States}
The premise of this paper is that if X states contain all of the same concurrence-purity (CP) combinations available to general states \textit{for each rank}, then by virtue of the fact that entanglement-preserving unitary (EPU) matrices preserve purity, concurrence, and rank of all states, then we should be able to transform any general state to an X state using a single EPU operation.

Therefore, first we must investigate all possible CP combinations available to \textit{general} states, and compare them to the MEMS of (\ref{eq:5}) to verify that they indeed represent a maximum for \textit{all} states, and not just X states.

Although it is popular to use the participation ratio $\frac{1}{\text{tr}(\rho^2)}$, we use the purity here since it is more ubiquitous in quantum information, and powers of it are directly proportional to $C$, rather than inversely proportional.

\begin{figure}[H]
\centering
\includegraphics[width=0.99\linewidth]{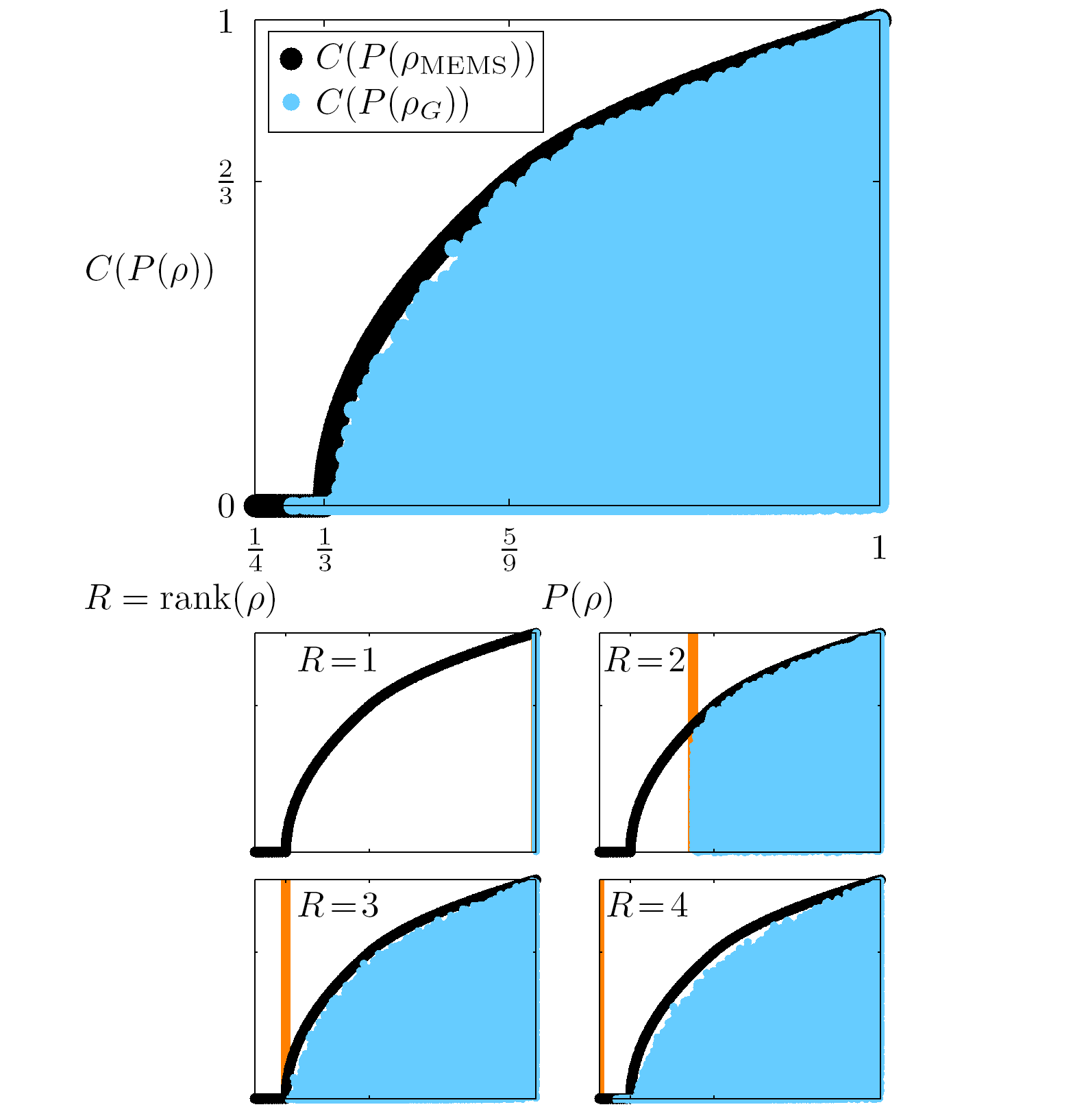}
\caption[]{\label{fig:1}(color online) Concurrence vs.~purity for maximally entangled mixed states $\rho _{\text{MEMS}}$ and $1,000,000$ general states $\rho_G$.  The lower plots are CP values of $250,000$ general states $\rho_G$ of rank $R$, with the minimal purity lines at $P_{\text{min}}=\frac{1}{R}$.}
\end{figure}

For CP relations, a more useful parameterization for $\rho _{\text{MEMS}}$ is in terms of purity $P$.  In fact, evaluating this reveals a \textit{third} case is necessary to ensure that a MEMS exists for every possible purity value, which yields
\begin{equation}
\rho _{\text{MEMS}}\!  \equiv\!\! \left\{\!\! {\begin{array}{*{20}l}
   {\left(\! \begin{array}{l}
 {\textstyle{{b_P  - 1} \over 2}}\left( {E_1  + E_2  + E_4 } \right) \\ 
  + {\textstyle{{5 - 3b_P } \over 2}}{\textstyle{1 \over 4}}I \\ 
 \end{array}\! \right)\!\!;} &\!\! {P \in [{\textstyle{1 \over 4}},{\textstyle{1 \over 3}}]}  \\
   {\left(\!\!\! \begin{array}{l}
 \sqrt {2(P - {\textstyle{1 \over 3}})} \rho _{\Phi ^ +  }  + {\textstyle{1 \over 3}}E_2  \\ 
  + \left({\textstyle{1 \over 3}}\! -\! \sqrt {{\textstyle{1 \over 2}}(P - {\textstyle{1 \over 3}})} \right)\!(E_1  + E_4 ) \\ 
 \end{array}\!\! \right)\!\!;} &\!\! {P \in [{\textstyle{1 \over 3}},{\textstyle{5 \over 9}}]}  \\
   {{\textstyle{{1 + \sqrt {2P - 1} } \over 2}}\rho _{\Phi ^ +  }  +\!\! \left(1 -\! {\textstyle{{1 + \sqrt {2P - 1} } \over 2}}\right)\! E_{2}; } &\!\! {P \in [{\textstyle{5 \over 9}},1]}  \\
\end{array}} \right.
\label{eq:7}
\end{equation}
where $b_P\equiv\sqrt {1 - {\textstyle{{16} \over 3}}(1 - 4P)}$, and $\rho _{\Phi ^ +  }  \equiv |\Phi ^ +  \rangle \langle \Phi ^ +  |$, $E_1  \equiv |0,0\rangle \langle 0,0|$, $E_2  \equiv |0,1\rangle \langle 0,1|$, and $E_4  \equiv |1,1\rangle \langle 1,1|$.  This then reveals the MEMS CP relation as
\begin{equation}
C(P(\rho _{\text{MEMS}} ))\! =\! \left\{\! {\begin{array}{*{20}l}
   0; & {P \in [{\textstyle{1 \over 4}},{\textstyle{1 \over 3}}]}  \\
   {\sqrt {2(P - {\textstyle{1 \over 3}})}; } & {P \in [{\textstyle{1 \over 3}},{\textstyle{5 \over 9}}]}  \\
   {{\textstyle{{1 + \sqrt {2P - 1} } \over 2}};} & {P \in [{\textstyle{5 \over 9}},1],}  \\
\end{array}} \right.
\label{eq:8}
\end{equation}
which was obtained by putting (\ref{eq:7}) into (\ref{eq:4}), and matches the solid curve at the top of Fig.~\ref{fig:1}.

As the top plot of Fig.~\ref{fig:1} shows, the CP plot of $\rho _{\text{MEMS}}$ does indeed appear to match the extreme upper bound of all general two-qubit states.  Of course, this is not a proof, but rather strong evidence, since the main sample is an even distribution over all ranks of $1,000,000$ random general states.  The bottom plots show that the CP values accessible to general states of rank $R$ are bounded only below at a purity-wall of $P_{\text{min}}=\frac{1}{R}$, with the ``separable ball'' having maximum purity $\frac{1}{n-1}=\frac{1}{3}$, \cite[]{Zycz}.

The need to look at rank-specific CP values is because a single unitary matrix cannot change the rank of a state.  Thus, to show EPU-equivalence, we need to show that for each rank, the X states access all the same CP values available to general states $\rho_G$ for that same rank.
The proof that $\rho _{\text{MEMS}}$ contain the maximum concurrence values for each possible purity value was given in \cite[]{Ishi} for states up to rank $3$, and checked numerically for rank $4$.  Therefore, since MEMS are \textit{also} X states, it is a \textit{fact} that X states contain the upper bounds of all CP values up to rank 3, and we shall tentatively accept the numerical results of the rank-$4$ case of Fig.~\ref{fig:1} as evidence in favor of the hypothesis that X states contain the \textit{upper bound} of CP values for \textit{all} general states and ranks.
\subsection{\label{sec:II.B}Evidence that X States Contain All CP Combinations}
Now that we have given evidence that X states contain the highest extreme CP values available to all states, we need to show that they contain all lower values, as well.  To accomplish this, we now define several parametric state families that will be useful in what follows.
\subsubsection{\label{sec:II.B.1}Bell States and $\theta$ States}
As a starting point, recall the Bell states,
\begin{equation}
\rho_{\Phi^{\pm}} \equiv \left( {\begin{array}{*{20}r}
   {{\textstyle{1 \over 2}}} & 0 & 0 & { \pm {\textstyle{1 \over 2}}}  \\
   0 & 0 & 0 & 0  \\
   0 & 0 & 0 & 0  \\
   { \pm {\textstyle{1 \over 2}}} & 0 & 0 & {{\textstyle{1 \over 2}}}  \\
\end{array}} \right)\!,\;\rho_{\Psi^{\pm}} \equiv \left( {\begin{array}{*{20}r}
   0 & 0 & 0 & 0  \\
   0 & {{\textstyle{1 \over 2}}} & { \pm {\textstyle{1 \over 2}}} & 0  \\
   0 & { \pm {\textstyle{1 \over 2}}} & {{\textstyle{1 \over 2}}} & 0  \\
   0 & 0 & 0 & 0  \\
\end{array}} \right)\!,
\label{eq:9}
\end{equation}
which may generalized by two-parameter pure states,
\begin{equation}
\begin{array}{*{20}l}
   {\rho _{\Phi (\theta ,\phi )} } &\!\! { \equiv \left( {\begin{array}{*{20}c}
   {c_\theta ^2 } & 0 & 0 & {{\textstyle{1 \over 2}}s_{2\theta } e^{ - i\phi } }  \\
   0 & 0 & 0 & 0  \\
   0 & 0 & 0 & 0  \\
   {{\textstyle{1 \over 2}}s_{2\theta } e^{i\phi } } & 0 & 0 & {s_\theta ^2 }  \\
\end{array}} \right)\!,}  \\
   {\rho _{\Psi (\theta ,\phi )} } &\!\! { \equiv \left( {\begin{array}{*{20}c}
   0 & 0 & 0 & 0  \\
   0 & {c_\theta ^2 } & {{\textstyle{1 \over 2}}s_{2\theta } e^{ - i\phi } } & 0  \\
   0 & {{\textstyle{1 \over 2}}s_{2\theta } e^{i\phi } } & {s_\theta ^2 } & 0  \\
   0 & 0 & 0 & 0  \\
\end{array}} \right)\!.}  \\
\end{array}
\label{eq:10}
\end{equation}
where $c_\theta \equiv \cos(\theta)$ and $s_\theta \equiv \sin(\theta)$, $\theta\in [0,\frac{\pi}{2}]$, and $\phi\in [0,2\pi)$.  These states are separable at $\theta=\{0,\frac{\pi}{2}\}$, and maximally entangled at $\theta=\frac{\pi}{4}$, where they are equal to Bell states when $\phi =\{0,\pi\}$.  Intermediate $\theta$ values cause varying degrees of entanglement, and hence we may call these $\theta$ states.

Some noteworthy features of $\theta$ states are that they are all X states, and they contain all possible maximally entangled pure X states (though keep in mind that maximally entangled non-X pure states exist as well), as well as the four separable standard basis states $|{0,0}\rangle\langle{0,0}|$, $|{0,1}\rangle\langle{0,1}|$, $|{1,0}\rangle\langle{1,0}|$, and $|{1,1}\rangle\langle{1,1}|$.  We can even define real-valued $\theta$ states as $\rho _{\Phi^{+} (\theta )}\equiv\rho _{\Phi (\theta ,0 )}$, $\rho _{\Phi^{-} (\theta )}\equiv\rho _{\Phi (\theta ,\pi )}$, $\rho _{\Psi^{+} (\theta )}\equiv\rho _{\Psi (\theta ,0 )}$, and $\rho _{\Psi^{-} (\theta )}\equiv\rho _{\Psi (\theta ,\pi )}$.

The purpose of defining these states is to parameterize all possible X states, which we will explore next.
\subsubsection{\label{sec:II.B.2}Generalized Two-Qubit X States}
Here, we wish to find a general form for all possible X states, so that we can search the full range of CP values accessible to X states.  To obtain a general X form, note that all X states have at most, two unique non-zero off-diagonal elements, $\rho_{3,2}$ and $\rho_{4,1}$. These are generally formed by convex sums of complex numbers.

Since any complex number can be represented as a vector on a complex plane, and since off-diagonal elements of pure states are the geometric mean of the corresponding diagonal elements, the minimum number of pure states required to decompose either of the off-diagonal X elements alone is two.  Therefore, since there are two off-diagonal X elements, we need a minimum of four states to decompose a general mixed X-state.

Thus, we define the most general mixed state as an $11$-parameter mixed state,
\begin{equation}
\tilde{\rho}_{X}\!  \equiv p_1 \rho _{\Phi (\theta _1 ,\phi _1 )}\!  + p_2 \rho _{\Phi (\theta _2 ,\phi _2 )}\!   + p_3 \rho _{\Psi (\theta _3 ,\phi _3 )}\!   + p_4 \rho _{\Psi (\theta _4 ,\phi _4 )}, 
\label{eq:11}
\end{equation}
where the probabilities have hyperspherical form
\begin{equation}
\begin{array}{*{20}l}
   {p_1 } & { \equiv c_{\vartheta _1 }^2 }  \\
   {p_2 } & { \equiv s_{\vartheta _1 }^2 c_{\vartheta _2 }^2 }  \\
   {p_3 } & { \equiv s_{\vartheta _1 }^2 s_{\vartheta _2 }^2 c_{\vartheta _3 }^2 }  \\
   {p_4 } & { \equiv s_{\vartheta _1 }^2 s_{\vartheta _2 }^2 s_{\vartheta _3 }^2 },  \\
\end{array}
\label{eq:12}
\end{equation}
with parameters on $\{ \vartheta _1 ,\vartheta _2 ,\vartheta _3 ,\theta _1 ,\theta _2 ,\theta _3 ,\theta _4 \}  \in [0,\frac{\pi }{2}]$ and $\{ \phi _1 ,\phi _2 ,\phi _3 ,\phi _4 \}  \in [0,2\pi )$.  All possible X states can be obtained with suitable choice of these $11$ parameters, making $\tilde{\rho}_{X}$ the most general X state parameterization.

We can remove unnecessary parameters by considering
\begin{equation}
\rho _X  \equiv p_1 \rho _{\Phi ^ +  (\theta _1 )}  + p_2 \rho _{\Phi  (\theta_{2},\phi_{2} )}  + p_3 \rho _{\Psi ^ +  (\theta _3 )}  + p_4 \rho _{\Psi   (\theta_{4},\phi_{4} )},
\label{eq:13}
\end{equation}
where the probabilities are still given by (\ref{eq:12}), but the number of parameters is only nine.  The reason for only setting the first and third phase angles to zero is that each consecutive pair of pure states in (\ref{eq:13}) share off-diagonal elements, and thus for their sum to roam all complex phase values, only one of them has to be complex.

While (\ref{eq:13}) is the essential parameterization of the most general X states, we will find it convenient to study the following \textit{real}, rank-specific X states,
\begin{equation}
\begin{array}{*{20}l}
   {\rho _{X_1 } } &\!\! { \equiv \rho _{\Phi ^ +  (\theta _1 )} }  \\
   {\rho _{X_2 } } &\!\! { \equiv p_1 \rho _{\Phi ^ +  (\theta _1 )}  + p_2 \rho _{\Psi ^ +  (\theta _3 )} }  \\
   {\rho _{X_3 } } &\!\! {\equiv p_1 \rho _{\Phi ^ +  (\theta _1 )}  + p_2 \rho _{\Phi ^ -  (\theta _2 )}  + p_3 \rho _{\Psi ^ +  (\theta _3 )} }  \\
   {\rho _{X_4 } } &\!\! { \equiv p_1 \rho _{\Phi ^ +  (\theta _1 )}  + p_2 \rho _{\Phi ^ -  (\theta _2 )}  + p_3 \rho _{\Psi ^ +  (\theta _3 )}  + p_4 \rho _{\Psi ^ -  (\theta _4 )}, }  \\
\end{array}
\label{eq:14}
\end{equation}
where we will take $\rho_{X_{R}}$ as our canonical \textit{minimal} X state parameterization, since it can be linked to all other X states of rank $R$ by diagonal unitary phase transformations.  These states each have only $2R-1$ parameters.  For each state in (\ref{eq:14}), it is assumed that $\sum_{j}p_{j}=1$, so the parameterizations of the probabilities of the first three are different than those of (\ref{eq:12}).  Note that in (\ref{eq:14}), it is implied that each of the probabilities shown must be nonzero to ensure the specified rank, and for the same reason, the $\theta_k$ must be chosen to ensure that $\rho_{\Phi^{+}(\theta _1 )}\neq \rho_{\Phi^{-}(\theta _2 )}$ and $\rho_{\Psi^{+}(\theta _3 )}\neq \rho_{\Psi^{-}(\theta _4 )}$.  Now we are ready to investigate the CP qualities of X states.
\subsubsection{\label{sec:II.B.3}Evidence of All CP Combinations in X States}
Here, we take advantage of the fact that $\rho_X$ contains all possible X states to see if the set of X states has access to the same CP region accessible to general states.  Since symbolic evaluation of the concurrence of $\rho_X$ as a function of purity is difficult even with the help of (\ref{eq:4}), we limit ourselves to a random sample of states by randomly choosing angles for $\rho_X$, as shown below in Fig.~\ref{fig:2}.

As Fig.~\ref{fig:2} shows, the X states $\rho_X$ appear to have access to all of the same CP values available to general states, as seen by comparing Fig.~\ref{fig:2} to Fig.~\ref{fig:1}.  More importantly, however, the rank-specific plots show that the X states match the CP values of general states $\rho_G$ by rank, as well.  Due to this correspondence, a single EPU matrix can convert any $\rho_G$ to an X state, which is one of the central claims of this paper.

Again, while this randomly generated sample is no proof that this is true, it provides strong evidence in favor of the hypothesis that X states access the same set of CP combinations that general states do.  Furthermore, since the rank-specific plots use the \textit{real-valued} X states of (\ref{eq:14}), Fig.~\ref{fig:2} shows that real-valued X states access all CP combinations as well.
\begin{figure}[H]
\centering
\includegraphics[width=0.99\linewidth]{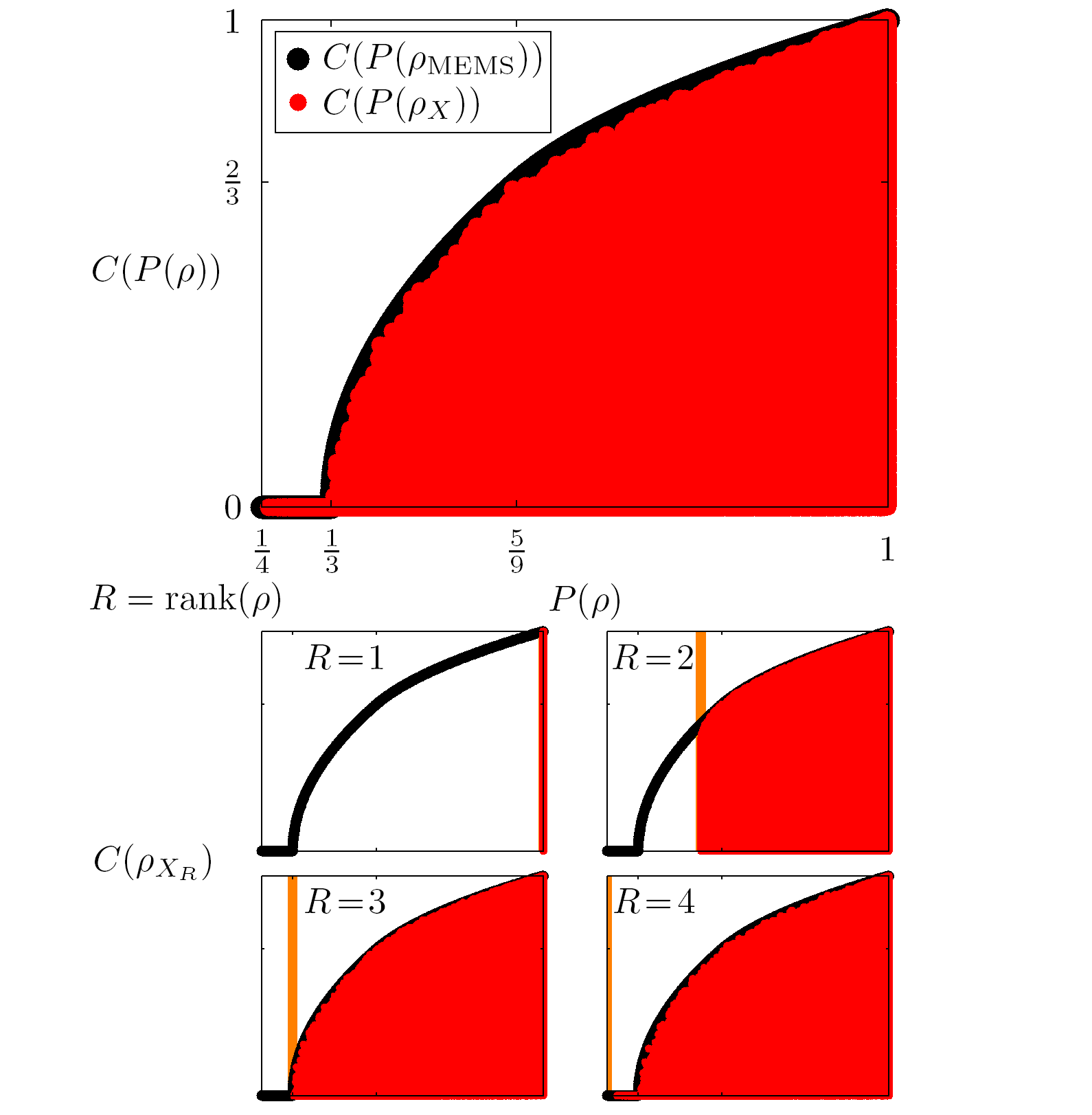}
\caption[]{\label{fig:2}(color online) Concurrence vs.~purity for maximally entangled mixed states $\rho _{\text{MEMS}}$ and $200,000$ general X states $\rho_X$.  The lower plots show $C(\rho_{X_{R}})$ by rank.  Comparison to Fig.~\ref{fig:1} indicates that X states of all ranks attain the same CP values as general states, implying EPU-equivalence.}
\end{figure}
\subsection{\label{sec:II.C}Convertibility of All States to X States}
\subsubsection{\label{sec:II.C.1}Finding the EPU X-Conversion Matrix If It Exists}
Suppose, as Fig.~\ref{fig:2} suggests, that X states access all CP values available to general states by rank.  Then, if that is true, here we prove the existence and form of a matrix that converts any general state into an X state.

First, if a general state $\rho_G$ and a \textit{particular} X state $\rho_X$ have the same purity and rank, then they can be related by a single unitary matrix.  Furthermore, if $\rho_X$ \textit{also} has the same \textit{concurrence} as $\rho_G$, then the unitary matrix that relates them also preserves the entanglement, which \textit{can} mean, but does \textit{not necessarily} mean it has a product form $U^{(1)}\otimes U^{(2)}$, though in general, it does not.  Also, since unitary matrices cannot change eigenvalues, $\rho_G$ and $\rho_X$ must have the same eigenvalues.

Therefore, expressing the states as $\rho _G  = \epsilon _{\rho _G } \Lambda\epsilon _{\rho _G }^{\dag} $ and $\rho _X  = \epsilon _{\rho _X } \Lambda\epsilon _{\rho _X }^{\dag}$, where $\epsilon_{\rho_G}$ and $\epsilon_{\rho_X}$ are eigenvector matrices whose columns are the eigenvectors of their respective states and $\Lambda$ is the diagonal matrix of eigenvalues, where we use the same for each in accordance with the above argument, then we can eliminate $\Lambda$ as
\begin{equation}
\Lambda = \epsilon _{\rho _G }^{\dag}  \rho _G \epsilon _{\rho _G }  = \epsilon _{\rho _X }^{\dag}  \rho _X \epsilon _{\rho _X },
\label{eq:15}
\end{equation}
from which we obtain the X-conversion transformation,
\begin{equation}
\Omega _X (\rho _G ) \equiv U\rho _G U^{\dag}   \equiv \epsilon _{\rho _X } \epsilon _{\rho _G }^{\dag}  \rho _G \epsilon _{\rho _G } \epsilon _{\rho _X }^{\dag}  = \rho _X ,
\label{eq:16}
\end{equation}
so the EPU matrix that transforms $\rho_G$ to $\rho_X$ is
\begin{equation}
U \equiv \epsilon _{\rho _X } \epsilon _{\rho _G }^{\dag} ,
\label{eq:17}
\end{equation}
and is entanglement-preserving since it cannot change the concurrence of $\rho_G$, which is true by the above definition that this particular $\rho_X$ has the same concurrence as $\rho_G$.  Therefore, (\ref{eq:17}) requires additional constraints to ensure entanglement preservation, which we will discuss soon.

Thus, we have shown that \textit{if} X states access all possible CP values for all ranks, then every general state $\rho_G$ is EPU-equivalent to an X state $\rho_X$ of the same concurrence, where the EPU matrix relating them is given by (\ref{eq:17}), provided that $C(\rho_{X})=C(\rho_{G})$.  See Sec.~\ref{sec:II.D.1} and App.~\ref{sec:AppA} for details about the form of such EPU matrices. 
\subsubsection{\label{sec:II.C.2}Demonstration that Arbitrary General States Can Be Transformed to X States}
Here we perform a simple test of the EPU-equivalence of general states with X states.  The test is as follows.

First, generate a random general state $\rho\equiv\rho_G$.  Then, measure its concurrence, purity, and rank.  Next, search a large number of real-valued rank-specific X states from (\ref{eq:14}) until finding one that yields $\rho'\equiv U\rho U^{\dag}$ with $C$ equal to that of $\rho$ within a certain tolerance, for the $U$ defined in (\ref{eq:17}).  Then, $\rho'$ is an X state with the same $C$ as general state $\rho$.  As a measure to quantify how close this new state is to X form, measure the sum of square-magnitudes of its anti-X unique off-diagonals by using $a(\rho')$, where
\begin{equation}
a(\rho)\equiv 4(|\rho_{2,1}|^{2}+|\rho_{3,1}|^{2}+|\rho_{4,2}|^{2}+|\rho_{4,3}|^{2}).
\label{eq:18}
\end{equation}
Since $a=0$ exactly only when all eight complex parts of the anti-X elements are zero, this can only happen for X states.  In the case when a state is the most ``non-X'' it can be, the sum of all anti-X-element magnitudes is $\frac{1}{4}$, in which case $a=1$.  Thus, $a$ is a good measure for how close a state is to an X state.
\begin{figure}[H]
\centering
\includegraphics[width=0.99\linewidth]{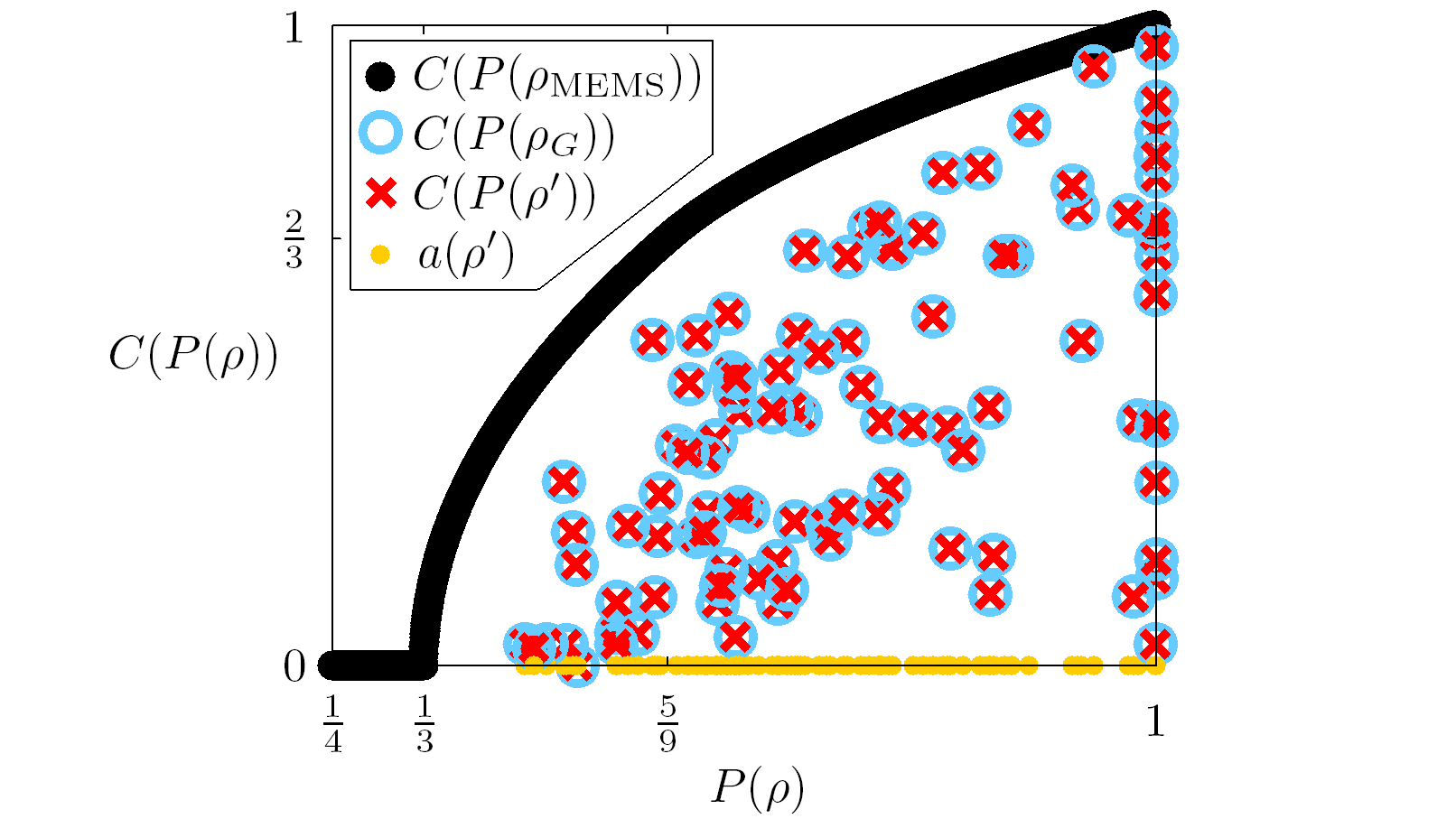}
\caption[]{\label{fig:3}(color online) Test to find the matrix $U\equiv\epsilon_{\rho_X}\epsilon_{\rho_G}^{\dag}$ to transform $100$ general states $\rho\equiv\rho_G$ to X states $\rho'\equiv U\rho U^{\dag}$ having $C(\rho')$ within $0.001$ of $C(\rho)$.  The CP values of input state $\rho$ are circles, while those of $\rho'$ are Xs.  The anti-X measure $a(\rho')$ is also plotted to show that $\rho'$ is truly an X state.  Not shown here, a total of $10,000$ consecutive successes have been obtained with this procedure.}
\end{figure}
As Fig.~\ref{fig:3} shows, the general states are all unitarily transformed to X states of the same concurrence to close approximation, and we may hypothesize that this can be done to any desired accuracy.  Note that the successes of this test were produced \textit{consecutively} using a while-loop for each input, meaning that \textit{no} failures were encountered in this sample of $10,000$ states, only the first $100$ of which are shown in Fig.~\ref{fig:3}.  Therefore, the above test provides compelling evidence to support the hypothesis that every general two-qubit state can be transformed to an X state with a single EPU transformation.  The fact that such a transformation is unitary is because $U$ is a product of unitary eigenvector matrices, and its entanglement-preservation is because $U$ also did not change the entanglement.  Thus, we have demonstrated strong evidence in favor of the main hypothesis of this paper.

As a more tangible example, given the random state,
\begin{equation}
%
\rho _G \! \approx\!\! \left(\! {\begin{array}{*{20}l}
   {0.31} & {0.12e^{ + i1.56} } & {0.23e^{ + i1.52} } & {0.28e^{ - i2.50} }  \\
   {0.12e^{ - i1.56} } & {0.07} & {0.09e^{+ i0.05} } & {0.12e^{+i2.27} }  \\
   {0.23e^{ - i1.52} } & {0.09e^{ - i0.05} } & {0.22} & {0.27e^{ +i2.13} }  \\
   {0.28e^{ + i2.50} } & {0.12e^{ - i2.27} } & {0.27e^{ - i2.13} } & {0.40}  \\
\end{array}}\!\! \right)\!\!,
\label{eq:19}
\end{equation}
the above procedure produces the transformed state
\begin{equation}
%
\rho ' \! \approx \! \left(\! {\begin{array}{*{20}l}
   {0.0285} & 0.0000 & 0.0000 & {0.0312}  \\
   0.0000 & {0.0960} & {0.2424} & 0.0000  \\
   0.0000 & {0.2424} & {0.8156} & 0.0000  \\
   {0.0312} & 0.0000 & 0.0000 & {0.0599}  \\
\end{array}}\! \right)\!,
\label{eq:20}
\end{equation}
where the anti-X elements in (\ref{eq:20}) are zeros to at least $15$ decimal places, and $C(\rho')\approx C(\rho_G)$ within $0.001$.

The prime importance of being able to find these kinds of transformations is that X states are always easier to work with, and can even enable symbolic computation of entanglement.  Therefore, the ability to transform any state to X form while preserving the original entanglement opens a door to symbolic computation of entanglement for \textit{any} input state.
\subsubsection{\label{sec:II.C.3}Simple Symbolic Example of EPU X Conversion}
In general, symbolic proof for EPU X conversions is very difficult, despite the ease with which numerical examples can be generated as in Fig.~\ref{fig:3}.  However, to illustrate the process with a simple case, we now look at an example simple enough to permit symbolic proof.

First, choose a non-X state from the restricted set,
\begin{equation}
\rho _G ;\;\;\left\{\! {\begin{array}{*{20}l}
   {C(\rho _G ) \in [0,1]}  \\
   {P(\rho _G ) \in [{\textstyle{1 \over 2}}(1 + C^2 ),1]}  \\
   {R(\rho _G ) \le 2,}  \\
\end{array}} \right.
\label{eq:21}
\end{equation}
where $R(\rho ) \equiv \text{rank}(\rho )$, and we shall abbreviate quantities of this ``general'' input state as $C \equiv C(\rho _G )$, $P \equiv P(\rho _G )$, and $R \equiv R(\rho _G )$.  Then, from purity and normalization, we know that $\lambda _1^2  + \lambda _2^2  = P$ and $\lambda _2  = 1 - \lambda _1$, which yields
\begin{equation}
\text{eig}(\rho _G ) = \{ {\textstyle{{1 + \sqrt {2P - 1} } \over 2}},{\textstyle{{1 - \sqrt {2P - 1} } \over 2}},0,0\}\equiv \{ \lambda _1 ,\lambda _2 ,\lambda _3 ,\lambda _4 \},
\label{eq:22}
\end{equation}
using descending-order convention.  Now, consider an X state of parametric concurrence $C \in [0,1]$ and purity $P \in [{\textstyle{1 \over 2}}(1 + C^2 ),1]$, defined as
\begin{equation}
\rho _X  \equiv \left( {\begin{array}{*{20}c}
   {{\textstyle{{1 + \sqrt {2P - 1 - C^2 } } \over 2}}} & 0 & 0 & {{\textstyle{1 \over 2}}C}  \\
   0 & 0 & 0 & 0  \\
   0 & 0 & 0 & 0  \\
   {{\textstyle{1 \over 2}}C} & 0 & 0 & {{\textstyle{{1 - \sqrt {2P - 1 - C^2 } } \over 2}}}  \\
\end{array}} \right)\!,
\label{eq:23}
\end{equation}
which has the properties that computing $\text{tr}(\rho _{X}^2)$ gives $P$, and using (\ref{eq:4}) yields $C$, and $C$ and $P$ are taken from $\rho_G$, to ensure that $\rho _{X}$ has the desired CP combination.  The secular equation of (\ref{eq:23}) yields its eigenvalues as
\begin{equation}
\text{eig}(\rho _X ) = \{ {\textstyle{{1 + \sqrt {2P - 1} } \over 2}},{\textstyle{{1 - \sqrt {2P - 1} } \over 2}},0,0\}\equiv \{ \xi _1 ,\xi _2 ,\xi _3 ,\xi _4 \},
\label{eq:24}
\end{equation}
which exactly match those of $\rho _G$ in (\ref{eq:22}).  Then, solving for the descending-order eigenvector matrix of $\rho _X$ gives
\begin{equation}
\epsilon _{\rho _X }  = \left(\! {\begin{array}{*{20}c}
   {{\textstyle{C \over {\sqrt {C^2  + (A - B)^2 } }}}} & {{\textstyle{C \over {\sqrt {C^2  + (A + B)^2 } }}}} & 0 & 0  \\
   0 & 0 & 0 & 1  \\
   0 & 0 & 1 & 0  \\
   {{\textstyle{{A - B} \over {\sqrt {C^2  + (A - B)^2 } }}}} & {{\textstyle{{ - (A + B)} \over {\sqrt {C^2  + (A + B)^2 } }}}} & 0 & 0  \\
\end{array}} \right)\!,
\label{eq:25}
\end{equation}
where $A \equiv \sqrt {2P - 1}$ and $B \equiv \sqrt {2P - 1 -C^{2}}$, and the column vectors of $\epsilon _{\rho _X }$ were verified to satisfy the eigenvalue equations $\rho _X \mathbf{v}_k  = \xi _k \mathbf{v}_k$ for $k=1,2,3,4$, where the eigenvectors $\mathbf{v}_k$ are column vectors of $\epsilon _{\rho _X }$ such that $\epsilon _{\rho _X }  = (\!\begin{array}{*{20}c}
   {\mathbf{v}_1 } &\!\! {\mathbf{v}_2 } &\!\! {\mathbf{v}_3 } &\!\! {\mathbf{v}_4 }  \\
\end{array}\!)$.  Thus, we obtain the following eigenvalue relation between the two states,
\begin{equation}
\epsilon _{\rho _G } ^{\dag}  \rho _G \epsilon _{\rho _G }  =\! \left(\! {\begin{array}{*{20}c}
   {{\textstyle{{1 + \sqrt {2P - 1} } \over 2}}} &\! 0 &\! 0 &\! 0  \\
   0 &\! {{\textstyle{{1 - \sqrt {2P - 1} } \over 2}}} &\! 0 &\! 0  \\
   0 &\! 0 &\! 0 &\! 0  \\
   0 &\! 0 &\! 0 &\! 0  \\
\end{array}}\! \right)\! = \epsilon _{\rho _X } ^{\dag}  \rho _X \epsilon _{\rho _X }, 
\label{eq:26}
\end{equation}
which is precisely the relation from (\ref{eq:15}), that allowed us to derive the X transformation in (\ref{eq:17}), which is
\begin{equation}
U = \epsilon _{\rho _X } \epsilon _{\rho _G } ^{\dag},
\label{eq:27}
\end{equation}
where here, $\epsilon _{\rho _X }$ is given explicitly in (\ref{eq:25}), and $\epsilon _{\rho _G }$ is the eigenvector matrix of the general input state from (\ref{eq:21}).

Thus, the X transformation we seek is
\begin{equation}
\begin{array}{*{20}l}
   {U\rho _G U^{\dag}  } &\!\! { = \epsilon _{\rho _X } \epsilon _{\rho _G }^ {\dag}  \rho _G \epsilon _{\rho _G } \epsilon _{\rho _X }^{\dag}  }  \\
   {} &\!\! { = \epsilon _{\rho _X } \text{diag}\{ \lambda _1 ,\lambda _2 ,0,0\} \epsilon _{\rho _X }^{\dag}  }  \\
   {} &\!\! { = \epsilon _{\rho _X } \text{diag}\{ \xi _1 ,\xi _2 ,0,0\} \epsilon _{\rho _X }^{\dag}  }  \\
   {} &\!\! { =\! \left( {\begin{array}{*{20}c}
   {{\textstyle{{1 + \sqrt {2P - 1 - C^2 } } \over 2}}} & 0 & 0 & {{\textstyle{1 \over 2}}C}  \\
   0 & 0 & 0 & 0  \\
   0 & 0 & 0 & 0  \\
   {{\textstyle{1 \over 2}}C} & 0 & 0 & {{\textstyle{{1 - \sqrt {2P - 1 - C^2 } } \over 2}}}  \\
\end{array}} \right)\! = \rho _X .}  \\
\end{array}
\label{eq:28}
\end{equation}

Thus, we have proven that, for the subset of general states $\rho_G$ defined in (\ref{eq:21}), it is always possible to unitarily transform $\rho_G$ into $\rho_X$ while preserving $C$ exactly.

Now that we have seen a simple example, we are ready to talk about the conditions for constructing the most general EPU X transformation.
\subsection{\label{sec:II.D}Expression of General States using X States}
\subsubsection{\label{sec:II.D.1}Parameterizing General States with X States}
If the claim that all states are EPU-equivalent to X states is true, as is supported by the test in Sec.~\ref{sec:II.C.2}, then we should be able to parameterize all states in terms of an X state transformed by an EPU matrix.  Therefore, using a real-valued X state $\rho_{X_{\text{real}}}$, the most general two-qubit state is
\begin{equation}
\rho = U_{\text{EPU}_{X}} \rho_{X_{\text{real}}} U_{\text{EPU}_{X}}^{\dag},
\label{eq:29}
\end{equation}
where $\rho_{X_{\text{real}}}\equiv \rho_{X_{4}}$ from (\ref{eq:14}) and $U_{\text{EPU}_{X}}$ is the most general EPU operation on X states, characterized by
\begin{equation}
\begin{array}{*{20}l}
   {U_{\text{EPU}_{X} }  \equiv } &\!\! {(U_{(2,1)} U_{(3,1)} U_{(3,2)} U_{(4,1)} U_{(4,2)} U_{(4,3)})^{\dag}D^{\dag} ;}  \\
   {} &\!\! {\text{s.t.}\;\;E(U_{\text{EPU}_{X}} \rho_{X_{\text{real}}} U_{\text{EPU}_{X}}^{\dag})=E(\rho_{X_{\text{real}}}),}  \\
\end{array}
\label{eq:30}
\end{equation}
where $E(\rho)$ is any valid entanglement measure, and $U_{(x,y)}$ is a single-qubit unitary matrix with only one superposition angle and one relative phase angle, where the parenthetical subscripts indicate the subspace upon which $U_{(x,y)}$ acts by referencing its only nonzero off-diagonal element, and $D$ is a diagonal unitary matrix.

The entanglement of each successive single-qubit-transformed state, such as $U_{(2,1)}^{\dag}D^{\dag} \rho _{X_{\text{real}}}D U_{(2,1)}$, is generally \textit{not} equivalent to the final entanglement.  Specifically, each successive single-qubit transformation typically has a \textit{different} concurrence which can be higher or lower than that of $\rho_{X_{\text{real}}}$, even though the final state $\rho$ in (\ref{eq:29}) \textit{does} have the same concurrence as $\rho_{X_{\text{real}}}$.

In fact, using single-qubit factorization of the EPU matrices of the test in  Fig.~\ref{fig:3}, it has been verified that EPU matrices \textit{do} generally have nontrivial unitary rotations on all single-qubit subspaces, of the form in (\ref{eq:30}).  Thus, in general, as proved in App.~\ref{sec:AppA}, $U_{\text{EPU}_{X}}$ is \textit{nonlocal}.

Furthermore, as seen in Fig.~\ref{fig:2}, we only need \textit{real} X states $\rho_{X_{\text{real}}}$ to reach all CP values, and in general, $\rho_{X_{\text{real}}}$ has $7$ degrees of freedom (DOF), as seen in (\ref{eq:14}).  Then, to see whether $U_{\text{EPU}_{X}}$ could have enough DOF to upgrade $\rho_{X_{\text{real}}}$ to a fully general state, note that discarding global phase, $U_{\text{EPU}_{X}}$ intrinsically has $15$ DOF, $n^2 -n=12$ of which belong to all the $U_{(x,y)}$ collectively, while $n-1=3$ belong to $D^{\dag}$ with its global phase discarded.  However, the only DOF that matter are those that persist in the transformed \textit{state}.  Since $D^{\dag}$ is adjacent to the X state, the zeros of the X state reduce the $3$ DOF of $D^{\dag}$ to only $2$ DOF, so $D^{\dag} \rho _{X_{\text{real}}}D$ has $9$ DOF.  That leaves the $12$ variables of $\{U_{(x,y)}\}$ to contribute the necessary $6$ DOF still needed to produce the $15$ DOF of a general state.

While it is true that a local unitary matrix $U^{(1)}\otimes U^{(2)}$ contains up to $6$ DOF, the more general unitary form of (\ref{eq:30}) offers a greater range of possibilities for supplying the $6$ DOF, without limiting us to local transformations.  Since (\ref{eq:30}) describes \textit{the most} general way in which one could make \textit{any} unitary matrix, and since numerical tests confirm that all single-qubit rotations are generally present in EPU matrices, then (\ref{eq:30}) is the most general form for EPU matrices (though admittedly an explicit method to determine the angular parameters is still unknown).  See App.~\ref{sec:AppA} for more details.

The strong numerical evidence provided by Fig.~\ref{fig:3} suggests that \textit{every} general state $\rho_G$ can be unitarily related to an X state of the same concurrence as $\rho_G$.  Therefore, since $U_{\text{EPU}_{X}}$ is the most general such unitary matrix, then (\ref{eq:29}) is the most general way in which a general state $\rho\equiv\rho_G$ can be parameterized with an X state.
\subsubsection{\label{sec:II.D.2}Parametric Concurrence for General States}
Given that the most general two-qubit state can be parameterized as (\ref{eq:29}), then due to the fact that
\begin{equation}
C(U_{\text{EPU}_{X}}\rho_{X}U_{\text{EPU}_{X}}^{\dag})=C(\rho_{X}),
\label{eq:31}
\end{equation}
where $U_{\text{EPU}_{X}}$ is given in (\ref{eq:30}), the concurrence of any general state is then conveniently given by Yu and Eberly's explicit concurrence of its corresponding X state using (\ref{eq:4}).  Therefore, to obtain a \textit{general} two-qubit state with a given concurrence, purity, and rank (CPR), first find an X state with that CPR, and then the set of all general states of that same CPR combination is parametrically accessible by putting that X state into (\ref{eq:29}) and choosing a $U_{\text{EPU}_{X}}$.  In that way, the concurrence of a general state can be determined by that of an X state.
\subsection{\label{sec:II.E}Entanglement-Preserving Depolarization as a Universal Measure of Entanglement}
Recalling the \textit{conceptual definition} of MEMS as states that maximize the \textit{entanglement} (however it is measured) for any given purity, consider the following observation.

\textbf{Domain of Constant Entanglement Theorem}: \textit{For every general state $\rho_{G}$ with entanglement $E(\rho_{G})$, there exists a MEMS $\rho_{\text{MEMS}}$ and a pure state $\rho_{P}$ that have the same entanglement so that $E(\rho_{\text{MEMS}})=E(\rho_{G})=E(\rho_{P})$.  Thus, the domain of constant entanglement is a family of states whose members of minimum purity are MEMS and whose members of maximum purity are pure, all of which share the same entanglement.}

\begin{figure}[H]
\centering
\includegraphics[width=0.99\linewidth]{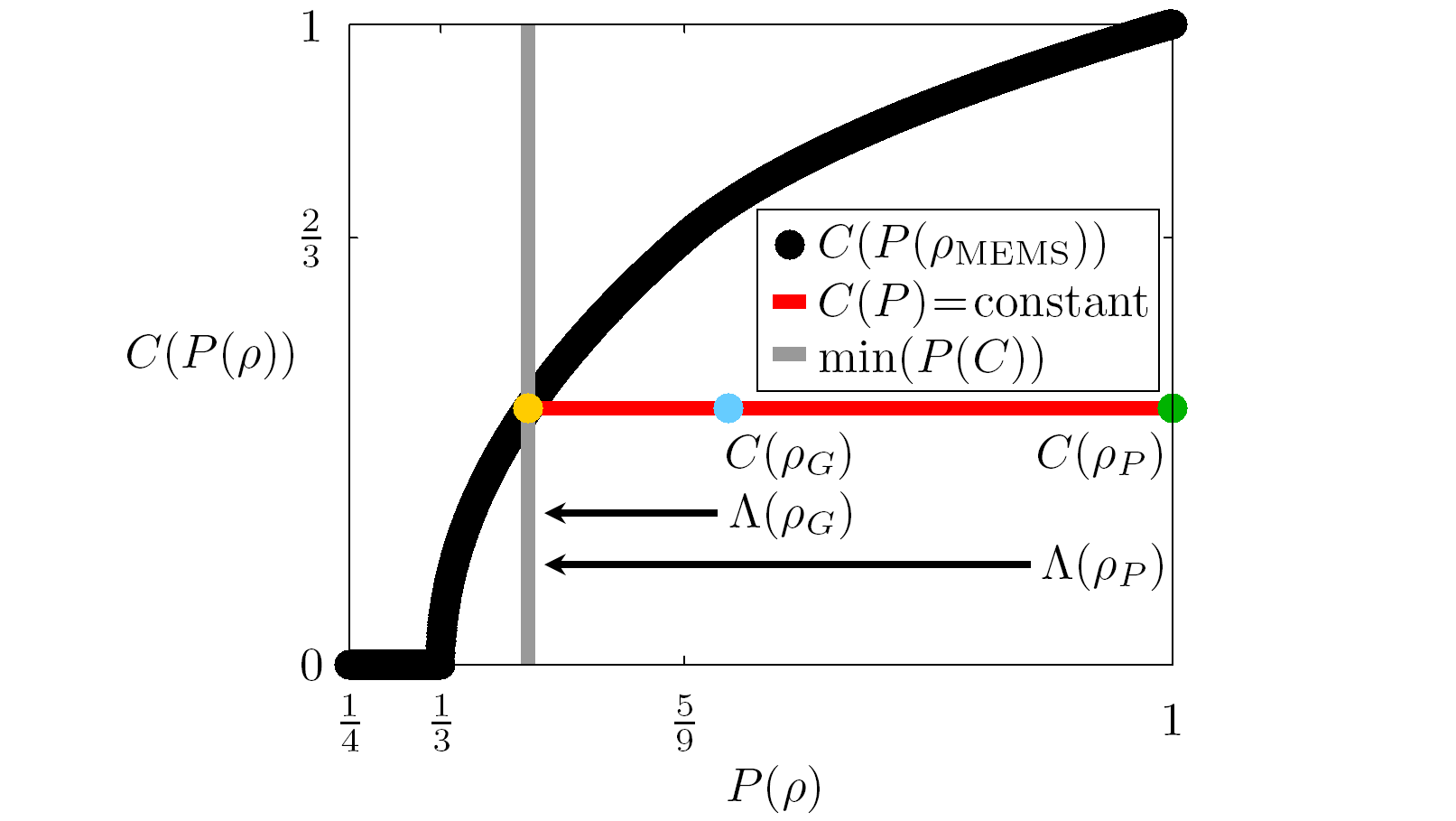}
\caption[]{\label{fig:4}(color online) Example showing that for any general state $\rho_{G}$, a line of constant concurrence passing through $C(\rho_{G})$ (blue) always contains both $C(\rho_{\text{MEMS}})$ (yellow) and the concurrence of a pure state $C(\rho_{P})$ (green).  Thus, in principle, if one could perform entanglement-preserving depolarizing transformations $\Lambda(\rho)$, then one could find the pure state of the same entanglement by showing that both $\rho_{G}$ and $\rho_{P}$ depolarized to the same minimum purity (the vertical gray line) for that constant entanglement.  The entanglement of $\rho_{G}$ could then be found from $\rho_{P}$ using any desired pure-state entanglement measure.  This could offer a way to compute entanglement in all systems, not merely $2\times 2$.}
\end{figure}

Note that the above observation is true regardless of the propositions we have made about X states.  Figure~\ref{fig:4} illustrates the essence of the Domain of Constant Entanglement Theorem (DCET) for CP values.

As Fig.~\ref{fig:4} shows, since all general states are bounded by MEMS and pure states, then for every general state $\rho_{G}$, a horizontal line of constant $C$ connects it to both a MEMS and a pure state of equal $C$.

Thus, the DCET suggests a new kind of entanglement measure.  Suppose there exists a transformation $\Lambda(\rho)$ that preserves entanglement while simultaneously depolarizing the input state as much as possible.  Since any depolarizing operation cannot increase the purity, that means $C(\Lambda(\rho))$ \textit{is always towards the left of $C(\rho)$}, unless $\rho$ is pure and maximally entangled.
Then, since $\Lambda(\rho)$ maximally depolarizes $\rho$ along the line of constant $C$, it will hit a \textit{purity wall} as shown by the vertical gray line in Fig.~\ref{fig:4}, which is imposed by the MEMS for that $C$.  For separable states, the purity wall is $\frac{1}{n}$, where the dimension $n=4$ for two qubits.

The method for finding entanglement of an arbitrary input state $\rho_G$ is then to first find $P(\Lambda(\rho_{G}))$, and then search a wide set of \textit{pure} states with a uniform distribution of entanglement values and find $P(\Lambda(\rho_{P}))$.  Finally,
\begin{equation}
\text{if}\;\;P(\Lambda (\rho _P )) = P(\Lambda (\rho _G )),\;\;\text{then}\;\;C(\rho _G ) = C(\rho _P ),
\label{eq:32}
\end{equation}
which says that if both $\rho_{G}$ and $\rho_{P}$ have the same minimal purity in a channel of entanglement-preserving depolarization $\Lambda(\rho)$, then the entanglement of the general state $\rho_{G}$ can be computed directly as the entanglement of the pure state $\rho_{P}$.

Since finding entanglement of pure states is always possible, and since  $\theta$ states allow us to parameterize this entanglement smoothly, then we can test an arbitrarily fine resolution of entanglement values for candidate pure states until finding the one whose minimal purity in $\Lambda$ matches that of the input state to any desired tolerance.

Thus, if $\Lambda(\rho)$ can be found in all multipartite systems, this offers a means of devising a universal entanglement measure.  Furthermore, note that this does \textit{not} require that we know anything about MEMS or even that the pure states have X form.

The difficulty is finding the entanglement-preserving depolarization channel $\Lambda(\rho)$.  One candidate is the generalization of the local-unitary rotation, the so-called doubly-stochastic local channel, defined as
\begin{equation}
L(\rho ) \equiv \sum\nolimits_k {p_k (U_k^{(1)}  \otimes U_k^{(2)} )} \rho (U_k^{(1)}  \otimes U_k^{(2)} )^{\dag},
\label{eq:33}
\end{equation}
where $\sum_{k}p_{k}=1$ and $p_{k}\in [0,1]\;\forall k$.  Unfortunately, such channels do not generally preserve entanglement.

Thus, at present, there is no known form of $\Lambda(\rho)$.  Yet, it is likely that the DCET holds true in all larger systems, as well.  The conceptual existence of $\Lambda(\rho)$ is an intriguing thought that prompts us to try to develop this new kind of constant-entanglement purification, where in this context we would not enlarge the system as with conventional purification, but rather merely find the pure state in the same system that has equal entanglement to an input state.  For now, we leave this as an open problem.
\subsubsection{\label{sec:II.E.1}States of Constant Entanglement}
As a possible aid to finding the entanglement-preserving depolarization channels $\Lambda(\rho)$, here we investigate a parametric family of states that allows us to precisely specify both concurrence $C$ and purity $P$.

First, consider the prototype for such states, given by
\begin{equation}
\rho _H ' \equiv\! \left\{\!\!\! {\begin{array}{*{20}l}
   {\left(\!\! \begin{array}{l}
 C|\Phi ^ +  \rangle \langle \Phi ^ +  | \\ 
  + (1 - C)\left(\! \begin{array}{l}
 (1 - p)E_2  \\ 
  + {\textstyle{1 \over 2}}p(E_1  + E_4 ) \\ 
 \end{array}\! \right) \\ 
 \end{array}\!\!\! \right)} & {p \in [0,1]}  \\
   {\left(\!\! \begin{array}{l}
 q|\Phi _{{\textstyle{1 \over 2}}\sin ^{ - 1} (C)}^ +  \rangle \langle \Phi _{{\textstyle{1 \over 2}}\sin ^{ - 1} (C)}^ +  | \\ 
  + (1 - q)\left(\! \begin{array}{l}
 C|\Phi ^ +  \rangle \langle \Phi ^ +  | \\ 
  + {\textstyle{1 \over 2}}(1 - C)(E_1  + E_4 ) \\ 
 \end{array}\! \right) \\ 
 \end{array}\!\!\! \right)} & {q \in [0,1],}  \\
\end{array}} \right.
\label{eq:34}
\end{equation}
where $E_1  \equiv |0,0\rangle \langle 0,0|, E_2  \equiv |0,1\rangle \langle 0,1|, E_3  \equiv |1,0\rangle \langle 1,0|, E_4  \equiv |1,1\rangle \langle 1,1|$, and we shall call these H states due to the fact that changes in only $p$ and $q$ cause the CP plot to be \textit{horizontal}, as desired.  Notice that the first term of the second state is a $\theta$ state with $\theta=\frac{1}{2}\sin^{-1}(C)$, and in this form, the two parts overlap.

The full CP parameterization of the H states is
\begin{equation}
\begin{array}{l}
 \rho _H (C,P) \equiv  \\ 
 \left\{\!\! {\begin{array}{*{20}l}
   {\rho _{H_{\text{I}} } } &\!\!\!\! {\equiv{\textstyle{{1 + b_P } \over 8}}(E_{1}+E_{2}+E_{4})+{\textstyle{{5 - 3b_P } \over 8}}E_{3}}  \\
   {\rho _{H_{\text{II}} } } &\!\!\!\! { \equiv\!\! \left(\!\! {\begin{array}{*{20}c}
   {{\textstyle{{2 + \sqrt {6P - 2 - 3C^2 } } \over 6}}} &\!\!\!\! 0 &\!\! 0 &\!\! {{\textstyle{1 \over 2}}C}  \\
   0 &\!\!\!\! {{\textstyle{{1 - \sqrt {6P - 2 - 3C^2 } } \over 3}}} &\!\! 0 &\!\! 0  \\
   0 &\!\!\!\! 0 &\!\! 0 &\!\! 0  \\
   {{\textstyle{1 \over 2}}C} &\!\!\!\! 0 &\!\! 0 &\!\! {{\textstyle{{2 + \sqrt {6P - 2 - 3C^2 } } \over 6}}}  \\
\end{array}}\!\!\! \right)}  \\
   {\rho _{H_{\text{III}} } } &\!\!\!\! { \equiv\!\! \left(\!\! {\begin{array}{*{20}c}
   {{\textstyle{{1 + \sqrt {2P - 1 - C^2 } } \over 2}}} & 0 & 0 & {{\textstyle{1 \over 2}}C}  \\
   0 & 0 & 0 & 0  \\
   0 & 0 & 0 & 0  \\
   {{\textstyle{1 \over 2}}C} & 0 & 0 & {{\textstyle{{1 - \sqrt {2P - 1 - C^2 } } \over 2}}}  \\
\end{array}}\!\! \right)\!\!,}  \\
\end{array}}\!\!\! \right. \\ 
 \end{array}
\label{eq:35}
\end{equation}
where $b_P  \equiv \sqrt {1 - {\textstyle{{16} \over 3}}\left( {1 - 4P} \right)}$, and to which the first case has been added to describe the region of purity below the separable cutoff.  These states are defined on intervals,
\begin{equation}
\left\{\!\! {\begin{array}{*{20}l}
   {\rho _{H_{\text{I}} } ;} &\! {C = 0,P \in [{\textstyle{1 \over 4}},{\textstyle{1 \over 3}})}  \\
   {\rho _{H_{\text{II}} } ;} &\!\!\!\!\! {\left\{\!\! \begin{array}{l}
 C \in [0,{\textstyle{2 \over 3}}),P \in [({\textstyle{1 \over 3}} + {\textstyle{1 \over 2}}C^2 ),{\textstyle{1 \over 2}}(1 + C^2 )) \\ 
 C \in [{\textstyle{2 \over 3}},1],P \in [{\textstyle{1 \over 2}}(1 + (2C - 1)^2 ),{\textstyle{1 \over 2}}(1 + C^2 )) \\ 
 \end{array} \right.}  \\
   {\rho _{H_{\text{III}} } ;} &\! {C \in [0,1],P \in [{\textstyle{1 \over 2}}(1 + C^2 ),1],}  \\
\end{array}}\!\! \right.
\label{eq:36}
\end{equation}
and Fig.~\ref{fig:5} plots the H states of (\ref{eq:35}) below.
\begin{figure}[H]
\centering
\includegraphics[width=0.99\linewidth]{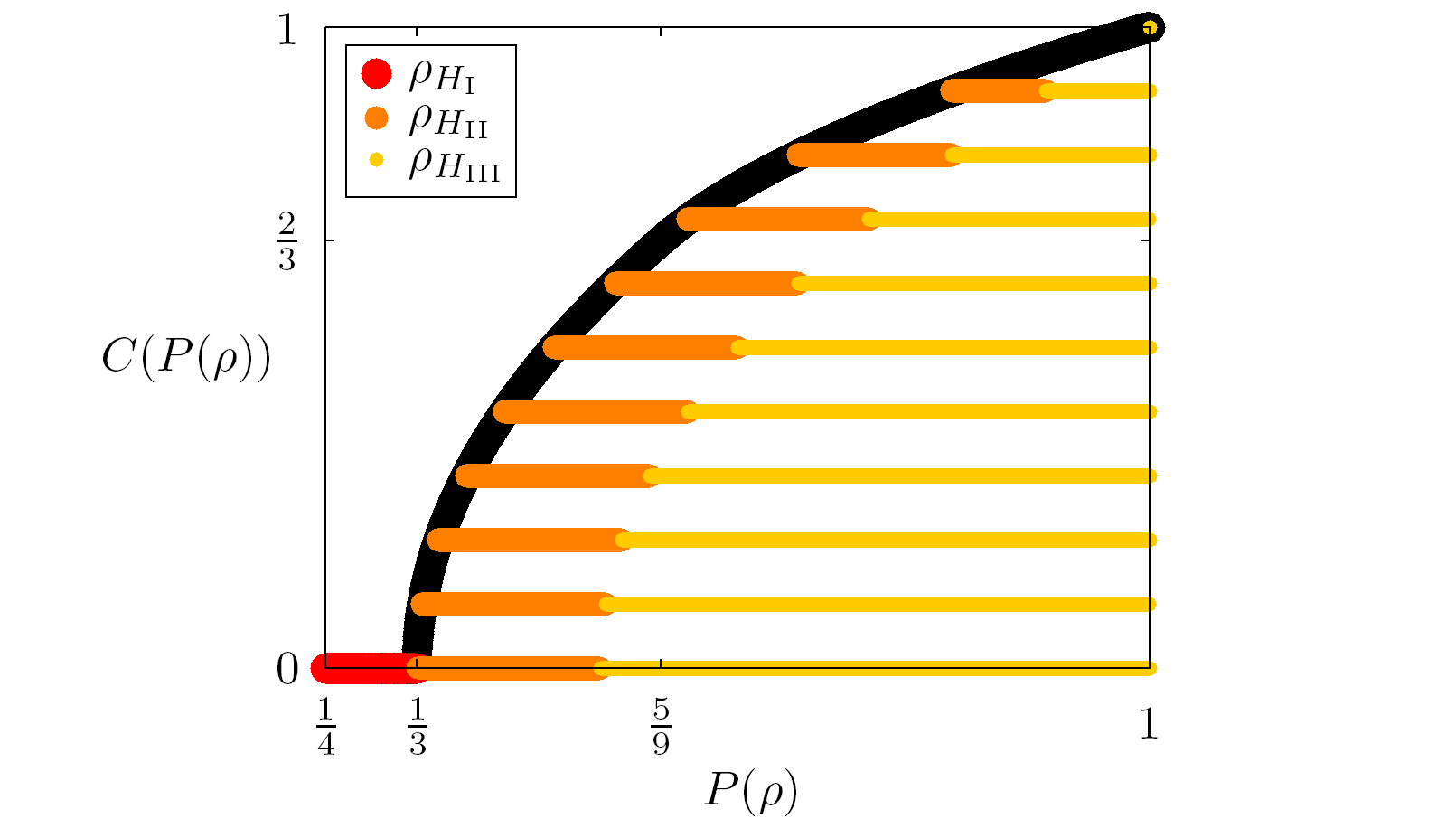}
\caption[]{\label{fig:5}(color online) Concurrence vs.~purity for the H states of (\ref{eq:35}) with constant $C$, against the MEMS in black.  Not shown is the case where holding $P$ constant and varying $C$ produces vertical CP lines.  The H states may aid in constructing entanglement-preserving depolarization channels.  Note that $C$ and $P$ can be given \textit{any} parameterization such as sine wave, step function, etc., but the horizontal form is most useful here, hence the name \textit{H states}.}
\end{figure}
The H states have the potential to be incredibly useful.  Since they span the full range of physical CP values, then any state EPU-equivalent to an H state can be linked directly to a $C$ value.  The procedure would be as follows.  First, from an input state $\rho_G$ of unknown $C$, measure $P$.  Then, holding $P$ constant, search an even distribution of different values of $C$ for a number of H states.  The one that that has the correct $C$ will have the same eigenvalues \textit{if the input state has the same rank as the H state}.  Unfortunately, the rank requirement makes this method only useful in \textit{some} cases.

At present, H states of rank $4$ have not yet been derived.  Alternatively, if there were a rank-changing transformation that preserved $C$, one could use that to adapt an input state to the rank of the H states.

If such a transformation could be found, and if H states could be found in all larger systems, then this would provide a method of obtaining a general entanglement measure.  Then, $C$ would be a parameter directly related to the superposition angle of the pure H states, and we could call such a measure \textit{the angle of entanglement}.

At present we leave this as an open challenge.  Now, we move to the last part of our discussion, which is the search for a generalized X form in larger systems that is also EPU-equivalent to all states.
\section{\label{sec:III}True-Generalized X States for Multipartite Systems}
Here, we generalize the idea of X states in all possible multipartite discrete systems by observing qualitative physical facts about the two-qubit case and applying those observations to larger systems.

The two main observations are firstly, that the \textit{anti-X} elements (the zeros in (\ref{eq:3})), are related to the partial trace operation, and secondly, that the X elements are identifiable by collectively observing a basis of maximally entangled states of a particularly simple form. 

We shall then show that these qualitative observations are equivalent and generalizable to all systems, working out several examples along the way, which will reveal that such states do not have a literal X shape.  Finally, we will examine the idea of the universality of such states by comparing these new \textit{true-generalized} X states (TGX states) with \textit{literal} X states.
\subsection{\label{sec:III.A}TGX Definition by Relation to Reductions}
\subsubsection{\label{sec:III.A.1}Relation of X Form to Partial Trace For Two Qubits}
Recall that the operation of partial tracing is tracing over only subsystems not to be retained while operating on the desired subsystems with the identity.  Thus, for two qubits, in terms of the full $\rho$, the reductions are
\begin{equation}
\begin{array}{*{20}l}
   {\rho ^{(1)}  =\! \left(\! {\begin{array}{*{20}c}
   {\rho _{1,1} \! +\! \rho _{2,2} } & {\rho _{1,3}  \! +\! \rho _{2,4} }  \\
   {\rho _{3,1}  \! +\! \rho _{4,2} } & {\rho _{3,3}  \! +\! \rho _{4,4} }  \\
\end{array}}\! \right)_{\;_{\;}}\!\!\!\!\!,}  \\
   {\rho ^{(2)}  =\! \left(\! {\begin{array}{*{20}c}
   {\rho _{1,1}  \! +\! \rho _{3,3} } & {\rho _{1,2}  \! +\! \rho _{3,4} }  \\
   {\rho _{2,1}  \! +\! \rho _{4,3} } & {\rho _{2,2}  \! +\! \rho _{4,4} }  \\
\end{array}}\! \right)^{\;}\!\!\!,}  \\
\end{array}
\label{eq:37}
\end{equation}
where superscripts refer to subsystem labels and are in parentheses to distinguish from matrix power notation.

Then, in complement to (\ref{eq:3}), define the anti-X matrix,
\begin{equation}
\rho _{\bar{X}}  \equiv \left( {\begin{array}{*{20}c}
   \cdot & {\rho _{1,2} } & {\rho _{1,3} } & \cdot  \\
   {\rho _{2,1} } & \cdot & \cdot & {\rho _{2,4} }  \\
   {\rho _{3,1} } & \cdot & \cdot & {\rho _{3,4} }  \\
   \cdot & {\rho _{4,2} } & {\rho _{4,3} } & \cdot  \\
\end{array}} \right)\!,
\label{eq:38}
\end{equation}
where the dots represent zeros.  Now, notice that the anti-X elements $\rho_{2,1},\rho_{3,1},\rho_{4,2},\rho_{4,3}$ all \textit{exclusively} form the unique off-diagonal elements of the reductions in (\ref{eq:37}).  

Therefore, one way to qualitatively form a hypothetical definition for X states is to define the anti-X elements as being those elements of the general density matrix that appear explicitly in the off-diagonal elements of all the reductions.  Then, \textit{an X state is any state for which the anti-X elements, as just defined, are all identically zero}.

Due to this last qualification, which arises by definition of the X form, one can see that the \textit{reductions of an X state will always be diagonal}, since elements of $\rho$ contributing to off-diagonal elements of reductions of X states are all identically zero.  However, it is important to note that this cannot be used in reverse.  Specifically, having diagonal reductions is merely \textit{necessary} for a state to have X form, but not sufficient.  This is because it is possible to have diagonal reductions where the anti-X components merely cancel each other out, but are not zero themselves.  An example of this is the maximally entangled state $|\psi \rangle  = {\textstyle{1 \over 2}}(|0,0\rangle  + |0,1\rangle  + i|1,0\rangle  - i|1,1\rangle )$, with density matrix
\begin{equation}
\rho  = {\textstyle{1 \over 4}}\left( {\begin{array}{*{20}r}
   1 & 1 & { - i} & i  \\
   1 & 1 & { - i} & i  \\
   i & i & 1 & { - 1}  \\
   { - i} & { - i} & { - 1} & 1  \\
\end{array}} \right),
\label{eq:39}
\end{equation}
which is clearly \textit{not} an X state, but nevertheless has diagonal reductions.

Thus, the requirement that full density elements appearing in off-diagonals of reductions be \textit{identically} zero is both necessary and sufficient for two-qubit X states, and we may express this definition operationally as
\begin{equation}
\rho _X  \equiv \rho  - \rho _{\bar X},
\label{eq:40}
\end{equation}
where the anti-X matrix $\rho _{\bar X}$ is defined as the matrix composed of all elements of the full $\rho$ appearing in the off-diagonals of all reductions, and zeros elsewhere.
\subsubsection{\label{sec:III.A.2}Generalization of X States by Partial Trace for All Possible Discrete Systems}
Here, we apply the proposed partial-trace definition to larger systems to see what that implies in those cases.  This is merely an extension of the qualitative observation of this fact for two qubits, and as such, requires no proof.  However, in terms of defining universal states, we will need to show that such states enjoy access to all possible entanglement-purity combinations by rank, as well.  

Since the only higher-dimensional case in which a necessary and sufficient entanglement measure is known is $2\times 3$, that is the only case for which we can test the universality of such states, and even then, the test is merely evidence, not proof.  

Therefore, starting with $2\times 3$, the reductions are
\begin{equation}
\begin{array}{l}
 \rho ^{(1)}  =\! \left(\! {\begin{array}{*{20}c}
   {\rho _{1,1}  \! +\! \rho _{2,2}  \! +\! \rho _{3,3} } & {\rho _{1,4}  \! +\! \rho _{2,5}  \! +\! \rho _{3,6} }  \\
   {\rho _{4,1}  \! +\! \rho _{5,2}  \! +\! \rho _{6,3} } & {\rho _{4,4}  \! +\! \rho _{5,5}  \! +\! \rho _{6,6} }  \\
\end{array}}\! \right)_{\;_{\;}}\!\!\!\!\!, \\ 
 \rho ^{(2)}  =\! \left(\! {\begin{array}{*{20}c}
   {\rho _{1,1}  \! +\! \rho _{4,4} } & {\rho _{1,2}  \! +\! \rho _{4,5} } & {\rho _{1,3}  \! +\! \rho _{4,6} }  \\
   {\rho _{2,1}  \! +\! \rho _{5,4} } & {\rho _{2,2}  \! +\! \rho _{5,5} } & {\rho _{2,3}  \! +\! \rho _{5,6} }  \\
   {\rho _{3,1}  \! +\! \rho _{6,4} } & {\rho _{3,2}  \! +\! \rho _{6,5} } & {\rho _{3,3}  \! +\! \rho _{6,6} }  \\
\end{array}}\! \right)^{\;}\!\!\!, \\ 
 \end{array}
\label{eq:41}
\end{equation}
and inspection of the off-diagonals reveals $\rho_{\bar{X}}$ to be
\begin{equation}
\rho _{\bar X}  = \left( {\begin{array}{*{20}c}
    \cdot  & {\rho _{1,2} } & {\rho _{1,3} } & {\rho _{1,4} } &  \cdot  &  \cdot   \\
   {\rho _{2,1} } &  \cdot  & {\rho _{2,3} } &  \cdot  & {\rho _{2,5} } &  \cdot   \\
   {\rho _{3,1} } & {\rho _{3,2} } &  \cdot  &  \cdot  &  \cdot  & {\rho _{3,6} }  \\
   {\rho _{4,1} } &  \cdot  &  \cdot  &  \cdot  & {\rho _{4,5} } & {\rho _{4,6} }  \\
    \cdot  & {\rho _{5,2} } &  \cdot  & {\rho _{5,4} } &  \cdot  & {\rho _{5,6} }  \\
    \cdot  &  \cdot  & {\rho _{6,3} } & {\rho _{6,4} } & {\rho _{6,5} } &  \cdot   \\
\end{array}} \right)\!.
\label{eq:42}
\end{equation}
Then, (\ref{eq:40}) yields the TGX state candidate for $2\times 3$ as
\begin{equation}
\rho _X  = \left( {\begin{array}{*{20}c}
   {\rho _{1,1} } &  \cdot  &  \cdot  &  \cdot  & {\rho _{1,5} } & {\rho _{1,6} }  \\
    \cdot  & {\rho _{2,2} } &  \cdot  & {\rho _{2,4} } &  \cdot  & {\rho _{2,6} }  \\
    \cdot  &  \cdot  & {\rho _{3,3} } & {\rho _{3,4} } & {\rho _{3,5} } &  \cdot   \\
    \cdot  & {\rho _{4,2} } & {\rho _{4,3} } & {\rho _{4,4} } &  \cdot  &  \cdot   \\
   {\rho _{5,1} } &  \cdot  & {\rho _{5,3} } &  \cdot  & {\rho _{5,5} } &  \cdot   \\
   {\rho _{6,1} } & {\rho _{6,2} } &  \cdot  &  \cdot  &  \cdot  & {\rho _{6,6} }  \\
\end{array}} \right)\!.
\label{eq:43}
\end{equation}
Immediately, we see that (\ref{eq:43}) does not exactly coincide with literal X states, due to the absence of $\rho_{5,2}$, and the presence of $\rho_{5,1}, \rho_{6,2},\rho_{4,2}$, and $\rho_{5,3}$.  We shall see further evidence that literal X states may not be the full story in the next section.

As another example, the reductions of $2\times 2\times 2$ are
\begin{equation}
\begin{array}{l}
 \rho ^{(1)}\!  =\! \left(\! {\begin{array}{*{20}c}
   {\rho _{1,1} \! +\! \rho _{2,2}  \! +\! \rho _{3,3}  \! +\! \rho _{4,4} } & {\rho _{1,5}  \! +\! \rho _{2,6}  \! +\! \rho _{3,7}  \! +\! \rho _{4,8} }  \\
   {\rho _{5,1}  \! +\! \rho _{6,2}  \! +\! \rho _{7,3}  \! +\! \rho _{8,4} } & {\rho _{5,5}  \! +\! \rho _{6,6}  \! +\! \rho _{7,7}  \! +\! \rho _{8,8} }  \\
\end{array}}\! \right)_{\;_{\;}}\!\!\!\!\!, \\ 
 \rho ^{(2)}\!  =\! \left(\! {\begin{array}{*{20}c}
   {\rho _{1,1}  \! +\! \rho _{2,2}  \! +\! \rho _{5,5}  \! +\! \rho _{6,6} } & {\rho _{1,3}  \! +\! \rho _{2,4}  \! +\! \rho _{5,7}  \! +\! \rho _{6,8} }  \\
   {\rho _{3,1}  \! +\! \rho _{4,2}  \! +\! \rho _{7,5}  \! +\! \rho _{8,6} } & {\rho _{3,3}  \! +\! \rho _{4,4}  \! +\! \rho _{7,7}  \! +\! \rho _{8,8} }  \\
\end{array}}\! \right)_{\;_{\;}}^{\;}\!\!\!\!\!, \\ 
 \rho ^{(3)}\!  =\! \left(\! {\begin{array}{*{20}c}
   {\rho _{1,1}  \! +\! \rho _{3,3}  \! +\! \rho _{5,5}  \! +\! \rho _{7,7} } & {\rho _{1,2}  \! +\! \rho _{3,4}  \! +\! \rho _{5,6}  \! +\! \rho _{7,8} }  \\
   {\rho _{2,1}  \! +\! \rho _{4,3}  \! +\! \rho _{6,5}  \! +\! \rho _{8,7} } & {\rho _{2,2}  \! +\! \rho _{4,4}  \! +\! \rho _{6,6}  \! +\! \rho _{8,8} }  \\
\end{array}}\! \right)^{\;}\!\!\!. \\ 
 \end{array}
\label{eq:44}
\end{equation}
Then, applying the definition in (\ref{eq:40}), we obtain the candidate TGX form for $2\times 2\times 2$ as
\begin{equation}
\rho _X  = \left( {\begin{array}{*{20}c}
   {\rho _{1,1} } &  \cdot  &  \cdot  & {\rho _{1,4} } &  \cdot  & {\rho _{1,6} } & {\rho _{1,7} } & {\rho _{1,8} }  \\
    \cdot  & {\rho _{2,2} } & {\rho _{2,3} } &  \cdot  & {\rho _{2,5} } &  \cdot  & {\rho _{2,7} } & {\rho _{2,8} }  \\
    \cdot  & {\rho _{3,2} } & {\rho _{3,3} } &  \cdot  & {\rho _{3,5} } & {\rho _{3,6} } &  \cdot  & {\rho _{3,8} }  \\
   {\rho _{4,1} } &  \cdot  &  \cdot  & {\rho _{4,4} } & {\rho _{4,5} } & {\rho _{4,6} } & {\rho _{4,7} } &  \cdot   \\
    \cdot  & {\rho _{5,2} } & {\rho _{5,3} } & {\rho _{5,4} } & {\rho _{5,5} } &  \cdot  &  \cdot  & {\rho _{5,8} }  \\
   {\rho _{6,1} } &  \cdot  & {\rho _{6,3} } & {\rho _{6,4} } &  \cdot  & {\rho _{6,6} } & {\rho _{6,7} } &  \cdot   \\
   {\rho _{7,1} } & {\rho _{7,2} } &  \cdot  & {\rho _{7,4} } &  \cdot  & {\rho _{7,6} } & {\rho _{7,7} } &  \cdot   \\
   {\rho _{8,1} } & {\rho _{8,2} } & {\rho _{8,3} } &  \cdot  & {\rho _{8,5} } &  \cdot  &  \cdot  & {\rho _{8,8} }  \\
\end{array}} \right)\!.
\label{eq:45}
\end{equation}
Notice that this includes a literal X shape as a \textit{subset}.  Therefore one question to investigate is whether this full form is needed to achieve all entanglement-purity combinations, or if literal X form is enough.

As a final example, which will reveal a special case in the next section, we consider $3\times 3$.  Its reductions are
\begin{equation}
\begin{array}{l}
 \rho ^{(1)}\!  =\!\! \left(\!\! {\begin{array}{*{20}c}
   {\rho _{1,1}  \!\!+\!\! \rho _{2,2}  \!\!+\!\! \rho _{3,3} } &\! {\rho _{1,4}  \!\!+\!\! \rho _{2,5}  \!\!+\!\! \rho _{3,6} } &\! {\rho _{1,7}  \!\!+\!\! \rho _{2,8}  \!\!+\!\! \rho _{3,9} }  \\
   {\rho _{4,1}  \!\!+\!\! \rho _{5,2}  \!\!+\!\! \rho _{6,3} } &\! {\rho _{4,4}  \!\!+\!\! \rho _{5,5}  \!\!+\!\! \rho _{6,6} } &\! {\rho _{4,7}  \!\!+\!\! \rho _{5,8}  \!\!+\!\! \rho _{6,9} }  \\
   {\rho _{7,1}  \!\!+\!\! \rho _{8,2} \!\!+\!\! \rho _{9,3} } &\! {\rho _{7,4}  \!\!+\!\! \rho _{8,5} \!\!+\!\! \rho _{9,6} } &\! {\rho _{7,7} \!\!+\!\! \rho _{8,8} \!\!+\!\! \rho _{9,9} }  \\
\end{array}}\!\! \right)_{\;_{\;}}\!\!\!\!\!, \\ 
 \rho ^{(2)}\!  =\!\! \left(\!\! {\begin{array}{*{20}c}
   {\rho _{1,1}  \!\!+\!\! \rho _{4,4}  \!\!+\!\! \rho _{7,7} } &\! {\rho _{1,2}  \!\!+\!\! \rho _{4,5} \!\!+\!\! \rho _{7,8} } &\! {\rho _{1,3}  \!\!+\!\! \rho _{4,6}  \!\!+\!\! \rho _{7,9} }  \\
   {\rho _{2,1}  \!\!+\!\! \rho _{5,4}  \!\!+\!\! \rho _{8,7} } &\! {\rho _{2,2}  \!\!+\!\! \rho _{5,5}  \!\!+\!\! \rho _{8,8} } &\! {\rho _{2,3}  \!\!+\!\! \rho _{5,6} \!\!+\!\! \rho _{8,9} }  \\
   {\rho _{3,1} \!\!+\!\! \rho _{6,4}  \!\!+\!\! \rho _{9,7} } &\! {\rho _{3,2}  \!\!+\!\! \rho _{6,5}  \!\!+\!\! \rho _{9,8} } &\! {\rho _{3,3}  \!\!+\!\! \rho _{6,6}  \!\!+\!\! \rho _{9,9} }  \\
\end{array}}\!\! \right)^{\;}\!\!\!, \\ 
 \end{array}
\label{eq:46}
\end{equation}
and then (\ref{eq:40}) yields the candidate $3\times 3$ TGX state as
\begin{equation}
\rho _X \!  =\! \left( {\begin{array}{*{20}c}
   {\rho _{1,1} } &  \cdot  &  \cdot  &  \cdot  & {\rho _{1,5} } & {\rho _{1,6} } &  \cdot  & {\rho _{1,8} } & {\rho _{1,9} }  \\
    \cdot  & {\rho _{2,2} } &  \cdot  & {\rho _{2,4} } &  \cdot  & {\rho _{2,6} } & {\rho _{2,7} } &  \cdot  & {\rho _{2,9} }  \\
    \cdot  &  \cdot  & {\rho _{3,3} } & {\rho _{3,4} } & {\rho _{3,5} } &  \cdot  & {\rho _{3,7} } & {\rho _{3,8} } &  \cdot   \\
    \cdot  & {\rho _{4,2} } & {\rho _{4,3} } & {\rho _{4,4} } &  \cdot  &  \cdot  &  \cdot  & {\rho _{4,8} } & {\rho _{4,9} }  \\
   {\rho _{5,1} } &  \cdot  & {\rho _{5,3} } &  \cdot  & {\rho _{5,5} } &  \cdot  & {\rho _{5,7} } &  \cdot  & {\rho _{5,9} }  \\
   {\rho _{6,1} } & {\rho _{6,2} } &  \cdot  &  \cdot  &  \cdot  & {\rho _{6,6} } & {\rho _{6,7} } & {\rho _{6,8} } &  \cdot   \\
    \cdot  & {\rho _{7,2} } & {\rho _{7,3} } &  \cdot  & {\rho _{7,5} } & {\rho _{7,6} } & {\rho _{7,7} } &  \cdot  &  \cdot   \\
   {\rho _{8,1} } &  \cdot  & {\rho _{8,3} } & {\rho _{8,4} } &  \cdot  & {\rho _{8,6} } &  \cdot  & {\rho _{8,8} } &  \cdot   \\
   {\rho _{9,1} } & {\rho _{9,2} } &  \cdot  & {\rho _{9,4} } & {\rho _{9,5} } &  \cdot  &  \cdot  &  \cdot  & {\rho _{9,9} }  \\
\end{array}} \right)\!,
\label{eq:47}
\end{equation}
which is another example where the definition of (\ref{eq:40}) does \textit{not} include the full literal X form as a subset.

Thus far, we have only looked at three examples of the proposed definition of TGX states from (\ref{eq:40}), but they illustrate that the definition is readily extended to arbitrary systems, regardless of whether or not it is useful, and they demonstrate various special cases.  Next, we look at a different proposal for defining TGX states, which we will see leads to the same results.
\subsection{\label{sec:III.B}TGX Definition by Identification of Simple Maximally Entangled Basis Sets}
Here, we propose to generalize a different qualitative fact from $2\times 2$. Recalling the Bell states of (\ref{eq:9}), notice \textit{that the union of their nonzero elements produces the general X state of (\ref{eq:3})}.

Therefore, since Bell states also form a compete basis and are maximally entangled, perhaps we can use such basis sets to define TGX states.  However, not just any maximally entangled basis sets would work.  As observed earlier, they must be composed of states for which elements contributing to off-diagonals of the reductions are identically zero.

Furthermore, this raises the question: are all systems guaranteed to have maximally entangled basis sets?  At present, there is no proof that this is true, but I encapsulate the idea in the following conjecture.

\textbf{Maximal Entanglement Basis Conjecture}: \textit{For all non-prime} dimensions $n\geq 4$, \textit{there exists at least one complete or overcomplete basis of maximally entangled pure states} $\{|\Phi _k \rangle \}$.

Then, supposing that the above conjecture (MEBC) is true, let an alternate definition for TGX states be \textit{the union of non-zero elements of all} simple \textit{maximally entangled basis sets, where a set is simple if its elements that contribute to off-diagonals of all single-site reductions are identically zero}.  In equation form,
\begin{equation}
\rho _X  \equiv \rho\circ\text{sgn}\left({\sum\nolimits_k {\text{abs}(\rho _{\Phi _k } ) }}\right),
\label{eq:48}
\end{equation}
which essentially means that for each possible simple maximally entangled state, if we find its element-wise absolute value, sum all such modified states, and take the sign, then performing the element-wise (Hadamard) product on a generic density matrix will yield TGX form.

Now we shall investigate this new definition for the three example systems of the previous section.  Note that we do not need to prove the MEBC in order to test this definition.  We are merely generalizing a fact that is true about two-qubit systems.  Thus, for $2\times 3$, one simple set of maximally entangled states is
\begin{equation}
\begin{array}{*{20}l}
   {|\Phi _1^ \pm  \rangle } &\!\! { \equiv {\textstyle{1 \over {\sqrt 2 }}}(|0,0\rangle } &\!\! { \pm |1,2\rangle ),}  \\
   {|\Phi _2^ \pm  \rangle } &\!\! { \equiv {\textstyle{1 \over {\sqrt 2 }}}(|0,1\rangle } &\!\! { \pm |1,0\rangle ),}  \\
   {|\Phi _3^ \pm  \rangle } &\!\! { \equiv {\textstyle{1 \over {\sqrt 2 }}}(|0,2\rangle } &\!\! { \pm |1,1\rangle ),}  \\
\end{array}
\label{eq:49}
\end{equation}
which is complete and orthonormal, and where we use the basis $\{|0,0\rangle,|0,1\rangle,|0,2\rangle,|1,0\rangle,|1,1\rangle,|1,2\rangle\}$.  Note that the reductions of these states are the most mixed it is possible for them to be in context of their origin in a $2\times 3$ system, but that $\rho^{(2)}$ will only be completely randomized in a \textit{2-dimensional subspace}, with one of its dimensions having a zero on the diagonal.

However, there is another \textit{different} simple complete maximally entangled basis (MEB) for $2\times 3$, which is
\begin{equation}
\begin{array}{*{20}l}
   {|\Psi _1^ \pm  \rangle } &\!\! { \equiv {\textstyle{1 \over {\sqrt 2 }}}(|0,0\rangle } &\!\! { \pm |1,1\rangle ),}  \\
   {|\Psi _2^ \pm  \rangle } &\!\! { \equiv {\textstyle{1 \over {\sqrt 2 }}}(|0,2\rangle } &\!\! { \pm |1,0\rangle ),}  \\
   {|\Psi _3^ \pm  \rangle } &\!\! { \equiv {\textstyle{1 \over {\sqrt 2 }}}(|0,1\rangle } &\!\! { \pm |1,2\rangle ).}  \\
\end{array}
\label{eq:50}
\end{equation}
Then, according to the definition in (\ref{eq:48}), the union of the nonzero elements of all possible MEBs (represented differently here to reveal the construction) is
\begin{equation}
\rho _X \! \sim\!\! \left(\! {\begin{array}{*{20}c}
   {\Phi _1 ,\Psi _1 } &  \cdot  &  \cdot  &  \cdot  & {\Psi _1 } & {\Phi _1 }  \\
    \cdot  & {\Phi _2 ,\Psi _3 } &  \cdot  & {\Phi _2 } &  \cdot  & {\Psi _3 }  \\
    \cdot  &  \cdot  & {\Phi _3 ,\Psi _2 } & {\Psi _2 } & {\Phi _3 } &  \cdot   \\
    \cdot  & {\Phi _2 } & {\Psi _2 } & {\Phi _2 ,\Psi _2 } &  \cdot  &  \cdot   \\
   {\Psi _1 } &  \cdot  & {\Phi _3 } &  \cdot  & {\Phi _3 ,\Psi _1 } &  \cdot   \\
   {\Phi _1 } & {\Psi _3 } &  \cdot  &  \cdot  &  \cdot  & {\Phi _1 ,\Psi _3 }  \\
\end{array}}\! \right)\!\!,
\label{eq:51}
\end{equation}
which agrees with the partial trace result from (\ref{eq:43}), and where the symbols shown indicate which states from (\ref{eq:49}) and (\ref{eq:50}) contributed to those elements.

In one respect, it is not surprising that this definition agrees with the partial trace definition, since both definitions require that the total form be built such that the elements contributing to off-diagonals in reductions are identically zero.  

However, this definition reveals several interesting things.  First, it is generally possible to have multiple MEBs that occupy generally \textit{different} matrix elements from each other, which is different from $2\times 2$ where the only simple MEB is the Bell basis.  Second, the total set of all possible simple MEBs appears to be enough to populate the entire TGX state as defined using the partial trace method.  Third, the union of all simple MEBs does not necessarily occupy all of the literal X state elements, as is apparent by the $0$ in the $\rho_{5,2}$ element in (\ref{eq:51}).

If a literal X form MEB were possible for $2\times 3$, then we would need to find a maximally entangled state that involved the $\rho_{5,2}$ and $\rho_{2,5}$ elements \textit{without} involving any of the elements not part of the TGX space already defined, since those elements are not part of literal X states.  But the only such states that could populate those elements are states composed of the second and fifth basis elements, such as $|\psi \rangle  = {\textstyle{1 \over {\sqrt 2 }}}(|0,1\rangle  + |1,1\rangle ) = {\textstyle{1 \over {\sqrt 2 }}}(|0\rangle  + |1\rangle ) \otimes |1\rangle $, which are clearly \textit{separable}.  Thus, one argument against literal X  form is that in the two-qubit case, the off-diagonal elements of X states can be exclusively associated with maximally entangled states, whereas for $2\times 3$, the literal X states can only be formed by including one separable state to get the coherence in the $\rho_{5,2}$ element.

While this alone is no reason to exclude the possibility of using literal X states, its departure from the qualitative form of two-qubit X states and the fact that TGX states exist that do not depart from that pattern suggest that literal X states may not always be the true generalization of the two-qubit case.

An example in which one of the simple MEBs \textit{does} form a literal X state is in $2\times 2\times 2$, for which the set
\begin{equation}
\begin{array}{*{20}l}
   {|\Phi _1^ \pm  \rangle } &\!\! { \equiv {\textstyle{1 \over {\sqrt 2 }}}(|0,0,0\rangle } &\!\!\! { \pm |1,1,1\rangle )}  \\
   {|\Phi _2^ \pm  \rangle } &\!\! { \equiv {\textstyle{1 \over {\sqrt 2 }}}(|0,0,1\rangle } &\!\!\! { \pm |1,1,0\rangle )}  \\
   {|\Phi _3^ \pm  \rangle } &\!\! { \equiv {\textstyle{1 \over {\sqrt 2 }}}(|0,1,0\rangle } &\!\!\! { \pm |1,0,1\rangle )}  \\
   {|\Phi _4^ \pm  \rangle } &\!\! { \equiv {\textstyle{1 \over {\sqrt 2 }}}(|0,1,1\rangle } &\!\!\! { \pm |1,0,0\rangle ),}  \\
\end{array}
\label{eq:52}
\end{equation}
populates the literal-X density elements given by
\begin{equation}
\rho  =\! \left( {\begin{array}{*{20}c}
   {\rho _{1,1} } &  \cdot  &  \cdot  &  \cdot  &  \cdot  &  \cdot  &  \cdot  & {\rho _{1,8} }  \\
    \cdot  & {\rho _{2,2} } &  \cdot  &  \cdot  &  \cdot  &  \cdot  & {\rho _{2,7} } &  \cdot   \\
    \cdot  &  \cdot  & {\rho _{3,4} } &  \cdot  &  \cdot  & {\rho _{3,6} } &  \cdot  &  \cdot   \\
    \cdot  &  \cdot  &  \cdot  & {\rho _{4,4} } & {\rho _{4,5} } &  \cdot  &  \cdot  &  \cdot   \\
    \cdot  &  \cdot  &  \cdot  & {\rho _{5,4} } & {\rho _{5,5} } &  \cdot  &  \cdot  &  \cdot   \\
    \cdot  &  \cdot  & {\rho _{6,3} } &  \cdot  &  \cdot  & {\rho _{6,6} } &  \cdot  &  \cdot   \\
    \cdot  & {\rho _{7,2} } &  \cdot  &  \cdot  &  \cdot  &  \cdot  & {\rho _{7,7} } &  \cdot   \\
   {\rho _{8,1} } &  \cdot  &  \cdot  &  \cdot  &  \cdot  &  \cdot  &  \cdot  & {\rho _{8,8} }  \\
\end{array}} \right)\!\!.
\label{eq:53}
\end{equation}
However, this is not the whole story, since $2\times 2\times 2$ \textit{also} admits another simple MEB given by
\begin{equation}
\begin{array}{*{20}l}
   {|\Phi _1^{ +  +  + } \rangle } &\!\! { \equiv {\textstyle{1 \over {\sqrt 4 }}}(|0,0,0\rangle } &\!\!\! { + |0,1,1\rangle } &\!\!\! { + |1,0,1\rangle } &\!\!\! { + |1,1,0\rangle )}  \\
   {|\Phi _1^{ -  -  + } \rangle } &\!\! { \equiv {\textstyle{1 \over {\sqrt 4 }}}(|0,0,0\rangle } &\!\!\! { - |0,1,1\rangle } &\!\!\! { - |1,0,1\rangle } &\!\!\! { + |1,1,0\rangle )}  \\
   {|\Phi _1^{ -  +  - } \rangle } &\!\! { \equiv {\textstyle{1 \over {\sqrt 4 }}}(|0,0,0\rangle } &\!\!\! { - |0,1,1\rangle } &\!\!\! { + |1,0,1\rangle } &\!\!\! { - |1,1,0\rangle )}  \\
   {|\Phi _1^{ +  -  - } \rangle } &\!\! { \equiv {\textstyle{1 \over {\sqrt 4 }}}(|0,0,0\rangle } &\!\!\! { + |0,1,1\rangle } &\!\!\! { - |1,0,1\rangle } &\!\!\! { - |1,1,0\rangle )}  \\
   {|\Phi _2^{ +  +  + } \rangle } &\!\! { \equiv {\textstyle{1 \over {\sqrt 4 }}}(|1,1,1\rangle } &\!\!\! { + |1,0,0\rangle } &\!\!\! { + |0,1,0\rangle } &\!\!\! { + |0,0,1\rangle )}  \\
   {|\Phi _2^{ -  -  + } \rangle } &\!\! { \equiv {\textstyle{1 \over {\sqrt 4 }}}(|1,1,1\rangle } &\!\!\! { - |1,0,0\rangle } &\!\!\! { - |0,1,0\rangle } &\!\!\! { + |0,0,1\rangle )}  \\
   {|\Phi _2^{ -  +  - } \rangle } &\!\! { \equiv {\textstyle{1 \over {\sqrt 4 }}}(|1,1,1\rangle } &\!\!\! { - |1,0,0\rangle } &\!\!\! { + |0,1,0\rangle } &\!\!\! { - |0,0,1\rangle )}  \\
   {|\Phi _2^{ +  -  - } \rangle } &\!\! { \equiv {\textstyle{1 \over {\sqrt 4 }}}(|1,1,1\rangle } &\!\!\! { + |1,0,0\rangle } &\!\!\! { - |0,1,0\rangle } &\!\!\! { - |0,0,1\rangle ),}  \\
\end{array}
\label{eq:54}
\end{equation}
which populates density elements,
\begin{equation}
\rho  =\!\! \left( {\begin{array}{*{20}c}
   {\rho _{1,1} } &  \cdot  &  \cdot  & {\rho _{1,4} } &  \cdot  & {\rho _{1,6} } & {\rho _{1,7} } &  \cdot   \\
    \cdot  & {\rho _{2,2} } & {\rho _{2,3} } &  \cdot  & {\rho _{2,5} } &  \cdot  &  \cdot  & {\rho _{2,8} }  \\
    \cdot  & {\rho _{3,2} } & {\rho _{3,3} } &  \cdot  & {\rho _{3,5} } &  \cdot  &  \cdot  & {\rho _{3,8} }  \\
   {\rho _{4,1} } &  \cdot  &  \cdot  & {\rho _{4,4} } &  \cdot  & {\rho _{4,6} } & {\rho _{4,7} } &  \cdot   \\
    \cdot  & {\rho _{5,2} } & {\rho _{5,3} } &  \cdot  & {\rho _{5,5} } &  \cdot  &  \cdot  & {\rho _{5,8} }  \\
   {\rho _{6,1} } &  \cdot  &  \cdot  & {\rho _{6,4} } &  \cdot  & {\rho _{6,6} } & {\rho _{6,7} } &  \cdot   \\
   {\rho _{7,1} } &  \cdot  &  \cdot  & {\rho _{7,4} } &  \cdot  & {\rho _{7,6} } & {\rho _{7,7} } &  \cdot   \\
    \cdot  & {\rho _{8,2} } & {\rho _{8,3} } &  \cdot  & {\rho _{8,5} } &  \cdot  &  \cdot  & {\rho _{8,8} }  \\
\end{array}} \right)\!\!,
\label{eq:55}
\end{equation}
which is clearly \textit{not} of literal X form.  The union of both (\ref{eq:53}) and (\ref{eq:55}) gives the MEB-motivated TGX form 
\begin{equation}
\rho _X \! =\!\! \left( {\begin{array}{*{20}c}
   {\rho _{1,1} } &  \cdot  &  \cdot  & {\rho _{1,4} } &  \cdot  & {\rho _{1,6} } & {\rho _{1,7} } & {\rho _{1,8} }  \\
    \cdot  & {\rho _{2,2} } & {\rho _{2,3} } &  \cdot  & {\rho _{2,5} } &  \cdot  & {\rho _{2,7} } & {\rho _{2,8} }  \\
    \cdot  & {\rho _{3,2} } & {\rho _{3,3} } &  \cdot  & {\rho _{3,5} } & {\rho _{3,6} } &  \cdot  & {\rho _{3,8} }  \\
   {\rho _{4,1} } &  \cdot  &  \cdot  & {\rho _{4,4} } & {\rho _{4,5} } & {\rho _{4,6} } & {\rho _{4,7} } &  \cdot   \\
    \cdot  & {\rho _{5,2} } & {\rho _{5,3} } & {\rho _{5,4} } & {\rho _{5,5} } &  \cdot  &  \cdot  & {\rho _{5,8} }  \\
   {\rho _{6,1} } &  \cdot  & {\rho _{6,3} } & {\rho _{6,4} } &  \cdot  & {\rho _{6,6} } & {\rho _{6,7} } &  \cdot   \\
   {\rho _{7,1} } & {\rho _{7,2} } &  \cdot  & {\rho _{7,4} } &  \cdot  & {\rho _{7,6} } & {\rho _{7,7} } &  \cdot   \\
   {\rho _{8,1} } & {\rho _{8,2} } & {\rho _{8,3} } &  \cdot  & {\rho _{8,5} } &  \cdot  &  \cdot  & {\rho _{8,8} }  \\
\end{array}} \right)\!,
\label{eq:56}
\end{equation}
which is precisely the form predicted by the partial-trace method in (\ref{eq:45}).  Thus, $2\times 2\times 2$ is an interesting example because it raises the question: if TGX states contain literal X states, are the literal X states enough to function as universal states, or is something lost by not using the full TGX states?

As a final example we see a new interesting special case arise for $3\times 3$.  Here, it does not seem possible to find a complete basis of simple maximally entangled states, but rather one can form an \textit{overcomplete} simple MEB by defining the states
\begin{equation}
\begin{array}{*{20}l}
   {|\Phi _1^{a,b} \rangle } &\!\! { \equiv {\textstyle{1 \over {\sqrt 3 }}}(|0,0\rangle } &\!\!\! { + a|1,1\rangle } &\!\!\! { + b|2,2\rangle) }  \\
   {|\Phi _2^{a,b} \rangle } &\!\! { \equiv {\textstyle{1 \over {\sqrt 3 }}}(|0,1\rangle } &\!\!\! { + a|1,2\rangle } &\!\!\! { + b|2,0\rangle) }  \\
   {|\Phi _3^{a,b} \rangle } &\!\! { \equiv {\textstyle{1 \over {\sqrt 3 }}}(|1,0\rangle } &\!\!\! { + a|2,1\rangle } &\!\!\! { + b|0,2\rangle) }  \\
   {|\Phi _4^{a,b} \rangle } &\!\! { \equiv {\textstyle{1 \over {\sqrt 3 }}}(|0,0\rangle } &\!\!\! { + a|1,2\rangle } &\!\!\! { + b|2,1\rangle) }  \\
   {|\Phi _5^{a,b} \rangle } &\!\! { \equiv {\textstyle{1 \over {\sqrt 3 }}}(|1,1\rangle } &\!\!\! { + a|0,2\rangle } &\!\!\! { + b|2,0\rangle) }  \\
   {|\Phi _6^{a,b} \rangle } &\!\! { \equiv {\textstyle{1 \over {\sqrt 3 }}}(|2,2\rangle } &\!\!\! { + a|0,1\rangle } &\!\!\! { + b|1,0\rangle), }  \\
\end{array}
\label{eq:57}
\end{equation}
where $a$ and $b$ are unit-magnitude relative phase factors.  The reason for including $a$ and $b$ is that (\ref{eq:57}) is neither complete nor overcomplete.  Apparently, only by using \textit{four} sets of (\ref{eq:57}), with $\{a,b\}=\{+1,+1\},\{+1,-1\},\{-1,+1\},\{-1,-1\}$, do we then get overcompleteness with $\frac{3}{8}\sum\nolimits_{\{a,b\}}\sum\nolimits_{k=1}^{6}{|\Phi_k^{a,b} \rangle\langle \Phi_k^{a,b}|}=I$.

Thus, the TGX form for $3\times 3$ appears to be
\begin{equation}
\rho _X \!  =\! \left( {\begin{array}{*{20}c}
   {\rho _{1,1} } &  \cdot  &  \cdot  &  \cdot  & {\rho _{1,5} } & {\rho _{1,6} } &  \cdot  & {\rho _{1,8} } & {\rho _{1,9} }  \\
    \cdot  & {\rho _{2,2} } &  \cdot  & {\rho _{2,4} } &  \cdot  & {\rho _{2,6} } & {\rho _{2,7} } &  \cdot  & {\rho _{2,9} }  \\
    \cdot  &  \cdot  & {\rho _{3,3} } & {\rho _{3,4} } & {\rho _{3,5} } &  \cdot  & {\rho _{3,7} } & {\rho _{3,8} } &  \cdot   \\
    \cdot  & {\rho _{4,2} } & {\rho _{4,3} } & {\rho _{4,4} } &  \cdot  &  \cdot  &  \cdot  & {\rho _{4,8} } & {\rho _{4,9} }  \\
   {\rho _{5,1} } &  \cdot  & {\rho _{5,3} } &  \cdot  & {\rho _{5,5} } &  \cdot  & {\rho _{5,7} } &  \cdot  & {\rho _{5,9} }  \\
   {\rho _{6,1} } & {\rho _{6,2} } &  \cdot  &  \cdot  &  \cdot  & {\rho _{6,6} } & {\rho _{6,7} } & {\rho _{6,8} } &  \cdot   \\
    \cdot  & {\rho _{7,2} } & {\rho _{7,3} } &  \cdot  & {\rho _{7,5} } & {\rho _{7,6} } & {\rho _{7,7} } &  \cdot  &  \cdot   \\
   {\rho _{8,1} } &  \cdot  & {\rho _{8,3} } & {\rho _{8,4} } &  \cdot  & {\rho _{8,6} } &  \cdot  & {\rho _{8,8} } &  \cdot   \\
   {\rho _{9,1} } & {\rho _{9,2} } &  \cdot  & {\rho _{9,4} } & {\rho _{9,5} } &  \cdot  &  \cdot  &  \cdot  & {\rho _{9,9} }  \\
\end{array}} \right)\!,
\label{eq:58}
\end{equation}
which we can actually find by only choosing one of the four sets of $\{a,b\}$, and which agrees exactly with the partial-trace result in (\ref{eq:47}).

Thus, we have seen three examples that support the equivalence between the partial-trace definition and the MEB definition for identifying TGX states.  These examples illustrate that literal X form is not always included in these TGX definitions, and that multiple TGX form MEB families are possible, and furthermore one may need to resort to overcomplete MEBs to find TGX states, although the partial-trace method is a more methodical and reliable means of identifying such states.

Now that we have raised a number of interesting questions, we are ready to take a deeper look at $2\times 3$ to see what can be learned from TGX states.
\subsection{\label{sec:III.C}Example in a Qubit-Qutrit System}
Since Peres' positive partial transpose (PPT) test provides a necessary and sufficient condition for separability in $2\times 3$, here we will briefly review the partial transpose and develop a useful entanglement measure from it.  Then, we will define MEMS for $2\times 3$ as well as a general parametric form for all mixed states in terms of both the TGX states and literal X states.  Then we will compare the entanglement-purity (EP) plots of these two families of states against the MEMS boundary to see if either or both are able to access all EP values.
\subsubsection{\label{sec:III.C.1}Partial Transpose and an Associated Entanglement Measure}
In the case of bipartite systems, the partial transpose (PT) operation can be conveniently expressed as
\begin{equation}
\begin{array}{l}
 \rho ^{T_1 }  \equiv \sum\limits_{a,b} {\left( {|a\rangle \langle b| \otimes I^{(2)} } \right)\rho \left( {|a\rangle \langle b| \otimes I^{(2)} } \right),}  \\ 
 \rho ^{T_2 }  \equiv \sum\limits_{a,b} {\left( {I^{(1)}  \otimes |a\rangle \langle b|} \right)\rho \left( {I^{(1)}  \otimes |a\rangle \langle b|} \right)},  \\ 
 \end{array}
\label{eq:59}
\end{equation}
where the basis kets $|a\rangle$ and $|b\rangle$ are the same as those in which the basis of $\rho$ is expressed for those subsystems, and $\rho ^{T_m }$ is the partial transpose with respect to subsystem $m$, while leaving the other subsystem alone.  Here, we use $\rho ^{T_1 }$, though both $\rho ^{T_1 }$ and $\rho ^{T_2 }$ have the same eigenvalues and are both Hermitian.

An intuitive way to visualize $\rho ^{T_1 }$ for $2\times 3$ is
\begin{equation}
\rho ^{T_1 }  = \left( {\begin{array}{*{20}c}
   {\begin{array}{*{20}c}
   {\rho _{1,1} } & {\rho _{1,2} } & {\rho _{1,3} }  \\
   {\rho _{2,1} } & {\rho _{2,2} } & {\rho _{2,3} }  \\
   {\rho _{3,1} } & {\rho _{3,2} } & {\rho _{3,3} }  \\
\end{array}} & \framebox[1.0\width]{$\begin{array}{*{20}c}
   {\rho _{4,1} } & {\rho _{4,2} } & {\rho _{4,3} }  \\
   {\rho _{5,1} } & {\rho _{5,2} } & {\rho _{5,3} }  \\
   {\rho _{6,1} } & {\rho _{6,2} } & {\rho _{6,3} }  \\
\end{array}$}  \\
   \framebox[1.0\width]{$\begin{array}{*{20}c}
   {\rho _{1,4} } & {\rho _{1,5} } & {\rho _{1,6} }  \\
   {\rho _{2,4} } & {\rho _{2,5} } & {\rho _{2,6} }  \\
   {\rho _{3,4} } & {\rho _{3,5} } & {\rho _{3,6} }  \\
\end{array}$} & {\begin{array}{*{20}c}
   {\rho _{4,4} } & {\rho _{4,5} } & {\rho _{4,6} }  \\
   {\rho _{5,4} } & {\rho _{5,5} } & {\rho _{5,6} }  \\
   {\rho _{6,4} } & {\rho _{6,5} } & {\rho _{6,6} }  \\
\end{array}}  \\
\end{array}} \right)\!,
\label{eq:60}
\end{equation}
where the boxes show regions of matrix elements of the original $\rho$ that have been exchanged by the PT operation.

In $2\times 3$, iff $\rho$ is separable, then $\rho ^{T_1 }$ is a physical state, meaning it has positive eigenvalues, regardless of whether $\rho$ is pure or mixed, discovered by Peres in \cite[]{Pere}.

When $\rho$ is entangled, at least one of the eigenvalues of $\rho ^{T_1 }$ are negative, yet they still sum to one.  Because of this, the sum of absolute values of the eigenvalues $||A||\equiv\text{tr}(\sqrt{A^{\dag}A})$, the Manhattan norm or $1$-norm, is an indicator of negativity of eigenvalues.  If they are all positive, $||\rho ^{T_1 }||=1$, but if any are negative, the fact that they still add to one will cause $||\rho ^{T_1 }||>1$.  Thus, a sensible entanglement measure based on the PPT test is
\begin{equation}
E_{T_m}(\rho ) \equiv {\textstyle{{||\rho ^{T_m } || - 1} \over {||\rho_{\text{ME}}^{T_m } || - 1}}},
\label{eq:61}
\end{equation}
where $\rho_{\text{ME}}$ is a maximally entangled state, and $m=1\;\text{or}\;2$, and we shall use $\rho ^{T_1 }$ here.  Note that (\ref{eq:61}) is just a rescaled negativity \cite[]{Vida}.  From (\ref{eq:49}), we find that $||\rho_{\text{ME}}^{T_1 } || - 1 = ||\rho_{\Phi_1^+}^{T_1} || - 1 = 1$, so in $2\times 3$, (\ref{eq:61}) is simply
\begin{equation}
E_{T_1}(\rho ) = ||\rho ^{T_1 } || - 1.
\label{eq:62}
\end{equation}
Now that (\ref{eq:62}) gives us a convenient entanglement measure for $2\times 3$, we need to identify the MEMS and show that they upper-bound $E_{T_1}(P(\rho) ) $ for all states.
\subsubsection{\label{sec:III.C.2}Identification of $2\times 3$ MEMS and Comparison to General States}
To this author's knowledge, MEMS for $2\times 3$, or for any system larger than $2\times 2$, have not been presented elsewhere.  This is partially due to the lack of general explicit entanglement measures in larger systems, and also to the inability to solve polynomial equations of order higher than $4$, which makes eigenvalue analysis very difficult.

Nevertheless, we can propose the $2\times 3$ MEMS as being those which are EPU-equivalent to
\begin{equation}
\rho _{\text{MEMS}}  \equiv\! \left\{\! {\begin{array}{*{20}l}
   \!\!{\left(\! \begin{array}{l}
 p|\Phi _1^ +  \rangle \langle \Phi _1^ +  | \\ 
  + {\textstyle{1 \over 5}}(1 + {\textstyle{p \over 2}})(E_2  + E_5 ) \\ 
  + {\textstyle{1 \over 5}}(1 - 2p)(E_1  + E_3  + E_6 ) \\ 
 \end{array}\! \right)\!\!;} & {p \in [0,{\textstyle{1 \over 2}}]}  \\
   {p|\Phi _1^ +  \rangle \langle \Phi _1^ +  | + {\textstyle{1 \over 2}}(1 - p)(E_2  + E_5 );} & {p \in [{\textstyle{1 \over 2}},1]}  \\
\end{array}} \right.
\label{eq:63}
\end{equation}
which are to be taken only as prototype states, and which were found partially by extending the two-qubit version of (\ref{eq:5}), and by observing the maximally entangled state presence in (\ref{eq:51}), and numerical experimentation.  Again here, $E_{1}\equiv|0,0\rangle\langle 0,0|$, $E_{2}\equiv|0,1\rangle\langle 0,1|$, $E_{3}\equiv|0,2\rangle\langle 0,2|$, $E_{4}\equiv|1,0\rangle\langle 1,0|$, $E_{5}\equiv|1,1\rangle\langle 1,1|$, $E_{6}\equiv|1,2\rangle\langle 1,2|$, and $|\Phi _1^ +  \rangle$ is given in (\ref{eq:49}).

Again, a more useful form is obtained by parameterizing for purity $P$, which reveals the missing first case,
\begin{equation}
\rho _{\text{MEMS}}  \equiv\! \left\{\!\!\! {\begin{array}{*{20}l}
   {\left(\!\! \begin{array}{l}
 {\textstyle{{f_P } \over 5}}\left( {E_1  + E_2  + E_3  + E_5  + E_6 } \right) \\ 
  + (1 - f_P ){\textstyle{1 \over 6}}I \\ 
 \end{array}\!\! \right)\!;} &\!\! {P\! \in [{\textstyle{1 \over 6}},{\textstyle{1 \over 5}}]}  \\
   {\left(\!\! \begin{array}{l}
 g_P \rho _{\Phi _1^ +  } \! + {\textstyle{1 \over 5}}(1 \!+\! {\textstyle{1 \over 2}}g_P )(E_2  + E_5 ) \\ 
  + {\textstyle{1 \over 5}}(1 - 2g_P )(E_1  + E_3  + E_6 ) \\ 
 \end{array}\!\! \right)\!;} &\!\! {P\! \in [{\textstyle{1 \over 5}},{\textstyle{3 \over 8}}]}  \\
   {{\textstyle{{1 + h_P } \over 3}}\rho _{\Phi _1^ +  }\!  + {\textstyle{1 \over 2}}(1 - {\textstyle{{1 + h_P } \over 3}})(E_2  + E_5 );} &\!\! {P\! \in [{\textstyle{3 \over 8}},1]}  \\
\end{array}}\!\!\! \right.
\label{eq:64}
\end{equation}
where $f_P  \equiv \sqrt {30(P - 1/6)} $, $g_P  \equiv \sqrt {(10/7)(P - 1/5)} $, and $h_P  \equiv \sqrt {6(P - 1/3)} $.

\begin{figure}[H]
\centering
\includegraphics[width=0.99\linewidth]{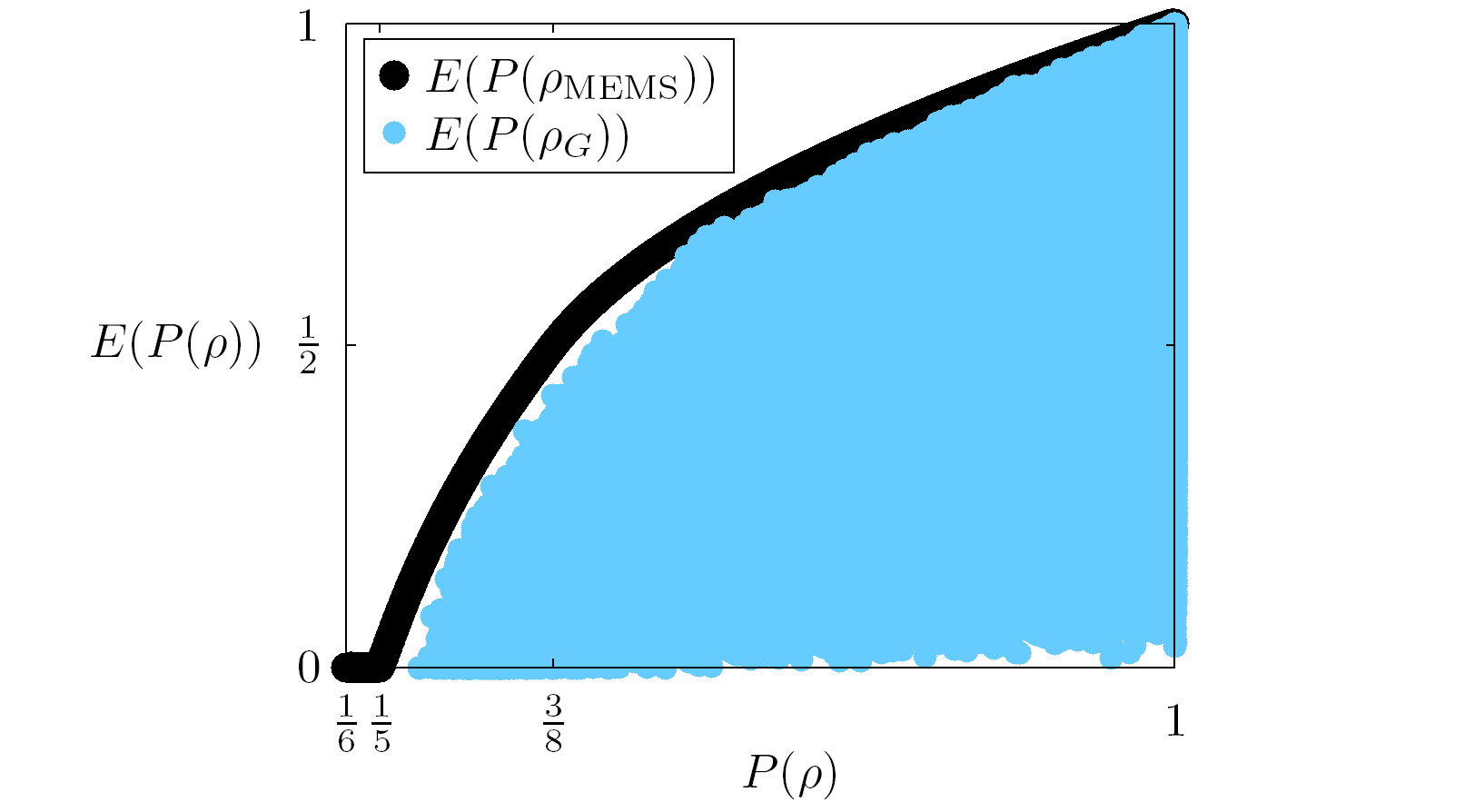}
\caption[]{\label{fig:6}(color online) Entanglement vs.~purity for the $2\times 3$ MEMS-candidates of (\ref{eq:64}) and $1,000,000$ random general states $\rho_G$, where $E(P(\rho))\equiv E_{T_{1}}(\rho)$ from (\ref{eq:62}).}
\end{figure}

Since these MEMS-candidates are merely heuristic observation of what works, Fig.~\ref{fig:6} offers numerical evidence that at least does not contradict the hypothesis that these are the correct MEMS.

Thus, we see that the candidate states of (\ref{eq:64}) do indeed have higher entanglement than a large sample of random general states for all purities.  Again this is no proof that these are the true MEMS, but if not, then they are near-MEMS, and will suffice for our purposes.  Now we move to the main goal of this section, which is to compare literal X states with TGX states.
\subsection{\label{sec:III.D}Literal X States vs.~True-Generalized X States}
First, we will consider \textit{literal} X states (LX states).  In fact, since these share some common members with the true-general X states (TGX states), the only states unique to LX states here are
\begin{equation}
\rho _{\text{L}_2 (\theta ,\phi )}  \equiv \left( {\begin{array}{*{20}c}
    \cdot  &  \cdot  &  \cdot  &  \cdot  &  \cdot  &  \cdot   \\
    \cdot  & {c_\theta ^2 } &  \cdot  &  \cdot  & {{\textstyle{1 \over 2}}s_{2\theta } e^{ - i\phi } } &  \cdot   \\
    \cdot  &  \cdot  &  \cdot  &  \cdot  &  \cdot  &  \cdot   \\
    \cdot  &  \cdot  &  \cdot  &  \cdot  &  \cdot  &  \cdot   \\
    \cdot  & {{\textstyle{1 \over 2}}s_{2\theta } e^{i\phi } } &  \cdot  &  \cdot  & {s_\theta ^2 } &  \cdot   \\
    \cdot  &  \cdot  &  \cdot  &  \cdot  &  \cdot  &  \cdot   \\
\end{array}} \right),
\label{eq:65}
\end{equation}
which are actually \textit{separable} since they are composed from bases $|0,1\rangle$ and $|1,1\rangle$, and where $\theta\in[0,\frac{\pi}{2}]$ and $\phi\in [0,2\pi)$.  The remaining LX states are then $\rho _{\text{L}_1 (\theta ,\phi )}  \equiv \rho _{\Phi _1 (\theta ,\phi )}$ and $\rho _{\text{L}_3 (\theta ,\phi )}  \equiv \rho _{\Psi _2 (\theta ,\phi )} $, which come from (\ref{eq:49}) and (\ref{eq:50}).  Thus, taken together, $\{\rho _{\text{L}_1 (\theta ,\phi )},\rho _{\text{L}_2 (\theta ,\phi )},\rho _{\text{L}_3 (\theta ,\phi )}\}$ constitute a state of literal X shape.  It is also convenient to define real-valued versions as $\rho _{\text{L}_k^ +  (\theta )}  \equiv \rho _{\text{L}_k (\theta ,0)} $ and $\rho _{\text{L}_k^ -  (\theta )}  \equiv \rho _{\text{L}_k (\theta ,\pi )}$.

\begin{figure}[H]
\centering
\includegraphics[width=0.99\linewidth]{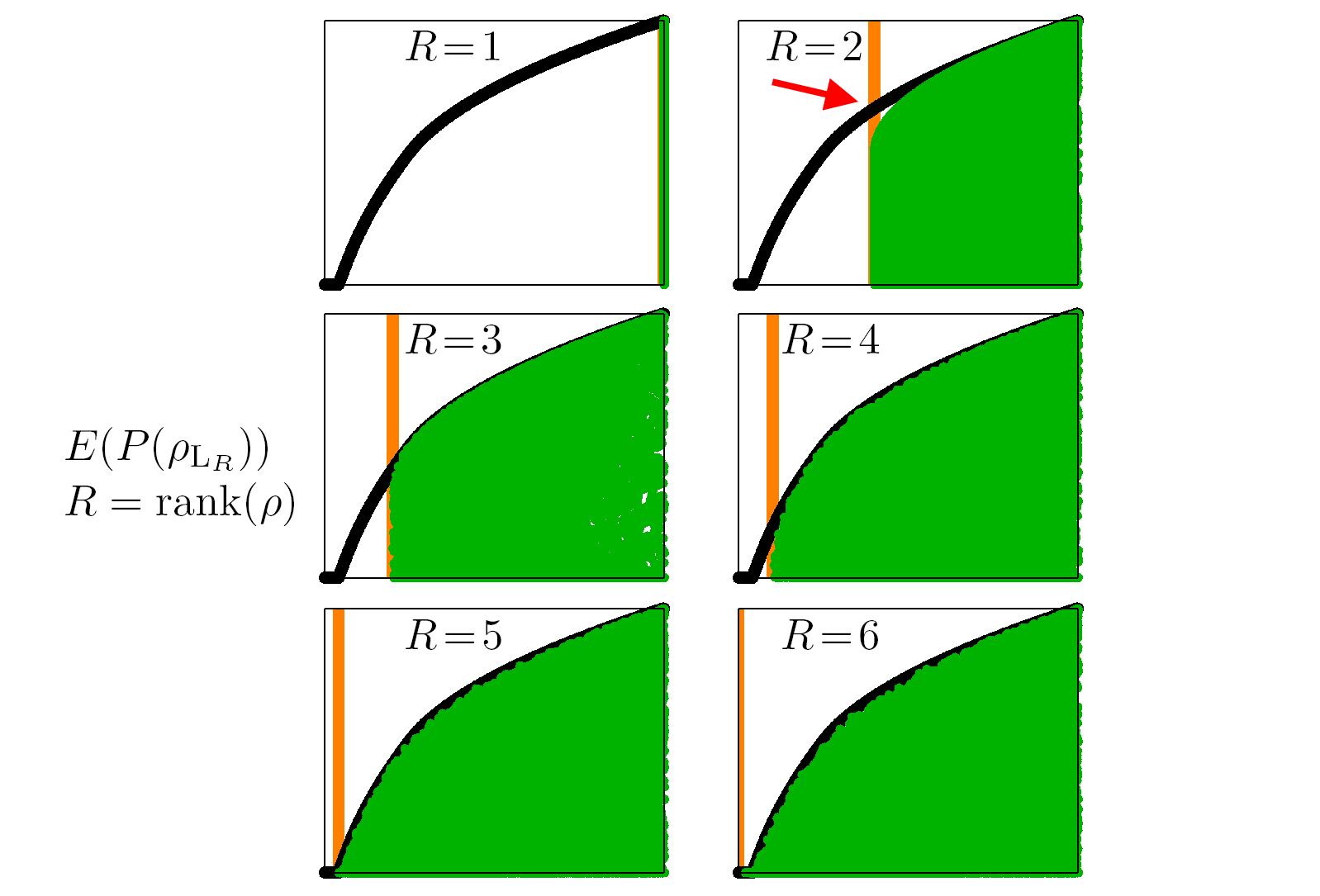}
\caption[]{\label{fig:7}(color online) Entanglement vs.~purity for $50,000$ literal X states $\rho_{L_R}$ of rank $R$ in $2\times 3$, each with the minimal purity line at $P_{\text{min}}=\frac{1}{R}$.  Notice that the rank-2 LX states are missing a small section of EP values in the upper-left corner, indicated by the red arrow.  Although not shown, the numerical values of each axis match those of Fig.~\ref{fig:6}.}
\end{figure}

Next, we need to define rank-specific LX states, as in (\ref{eq:14}).  However, since the Hilbert space is larger here, there are more options, and it appears that some combinations work better than others.  Therefore we shall choose the states found to get as close to the MEMS border as possible.  Thus, the rank-specific LX states are
\begin{equation}
\begin{array}{*{20}l}
   {\rho _{\text{LX}_1 } } &\!\! { \equiv \rho _{\text{L}_1^ +  (\theta _1 )} }  \\
   {\rho _{\text{LX}_2 } } &\!\! { \equiv p_1 \rho _{\text{L}_1^ +  (\theta _1 )}  + p_2 \rho _{\text{L}_2^ +  (\theta _2 )} }  \\
   {\rho _{\text{LX}_3 } } &\!\! { \equiv p_1 \rho _{\text{L}_1^ +  (\theta _1 )}  + p_2 \rho _{\text{L}_2^ +  (\theta _2 )}  + p_3 \rho _{\text{L}_2^ -  (\theta _3 )} }  \\
   {\rho _{\text{LX}_4 } } &\!\! { \equiv p_1 \rho _{\text{L}_1^ +  (\theta _1 )}  + p_2 \rho _{\text{L}_1^ -  (\theta _2 )}  + p_3 \rho _{\text{L}_2^ +  (\theta _3 )}  + p_4 \rho _{\text{L}_2^ -  (\theta _4 )} }  \\
   {\rho _{\text{LX}_5 } } &\!\! { \equiv\! \left(\! \begin{array}{l}
 p_1 \rho _{\text{L}_1^ +  (\theta _1 )}  + p_2 \rho _{\text{L}_2^ +  (\theta _2 )}  + p_3 \rho _{\text{L}_2^ -  (\theta _3 )}  \\ 
  + p_4 \rho _{\text{L}_3^ +  (\theta _4 )}  + p_5 \rho _{\text{L}_3^ -  (\theta _5 )}  \\ 
 \end{array}\!\! \right)}  \\
   {\rho _{\text{LX}_6 } } &\!\! { \equiv\! \left(\! \begin{array}{l}
 p_1 \rho _{\text{L}_1^ +  (\theta _1 )}  + p_2 \rho _{\text{L}_1^ -  (\theta _2 )}  + p_3 \rho _{\text{L}_2^ +  (\theta _3 )}  \\ 
  + p_4 \rho _{\text{L}_2^ -  (\theta _4 )}  + p_5 \rho _{\text{L}_3^ +  (\theta _5 )}  + p_6 \rho _{\text{L}_3^ -  (\theta _6 )}  \\ 
 \end{array}\!\! \right)\!,}  \\
\end{array}
\label{eq:66}
\end{equation}
where note that again, the probabilities and angles are chosen such that the intended rank is achieved.  The EP values for (\ref{eq:66}) are plotted in Fig.~\ref{fig:7}.
Figure~\ref{fig:7} shows that the best that rank-2 LX states can do is \textit{not enough} to cover all EP values!  This may seem trivial, given that the higher-rank LX states \textit{do} reach those EP values, however, since a \textit{single} EPU matrix preserves rank, if these observations are generally true, then that means that in the set of general states $\rho_G$, \textit{there are some rank-2 states that cannot be transformed to LX states by any single EPU matrix}.

The other rank-2 LX states tested in convex-sum pairs were $\{ \rho _{\text{L}_1^ +(\theta_1)  } ,\rho _{\text{L}_2^ - (\theta_2) } \}$, $\{ \rho _{\text{L}_1^ + (\theta_1) } ,\rho _{\text{L}_3^ + (\theta_2) } \} $, $\{ \rho _{\text{L}_1^ + (\theta_1) } ,\rho _{\text{L}_3^ - (\theta_2) } \} $, $\{ \rho _{\text{L}_2^ + (\theta_1) } ,\rho _{\text{L}_3^ + (\theta_2) } \}$, $\{ \rho _{\text{L}_2^ + (\theta_1) } ,\rho _{\text{L}_3^ - (\theta_2) } \}$, $\{ \rho _{\text{L}_1^ + (\theta_1) } ,\rho _{\text{L}_1^ - (\theta_2) } \}$, $\{ \rho _{\text{L}_2^ + (\theta_1)  } ,\rho _{\text{L}_2^ - (\theta_2) } \}$, $\{ \rho _{\text{L}_3^ + (\theta_1) } ,\rho _{\text{L}_3^ - (\theta_2) } \} $, and all rank-2 convex-sum pairs of $\{ \rho _{\text{L}_1^ + (\theta_1) } ,E_k \}$, where $E_k$ are the product states defined earlier in (\ref{eq:63}).  In all cases, no states performed better than what is shown in Fig.~\ref{fig:7}, and most did worse.  Variations were also tried with random phases, but these did no better either.

Of course, there may be rank-changing, EP-preserving transformations that \textit{could} map all general rank-2 states to LX states, but such transformations would be \textit{probabilistic}, requiring convex sums of local unitary operations, whereas if a single EPU matrix could do the job, we would have a deterministic means of connecting general states to LX states.  Since it seems that LX states do not reach all EP values for all ranks, we may hypothesize that in general, LX states are not fully universal to the set of general states by rank-1 EPU transformations.

\begin{figure}[H]
\centering
\includegraphics[width=0.99\linewidth]{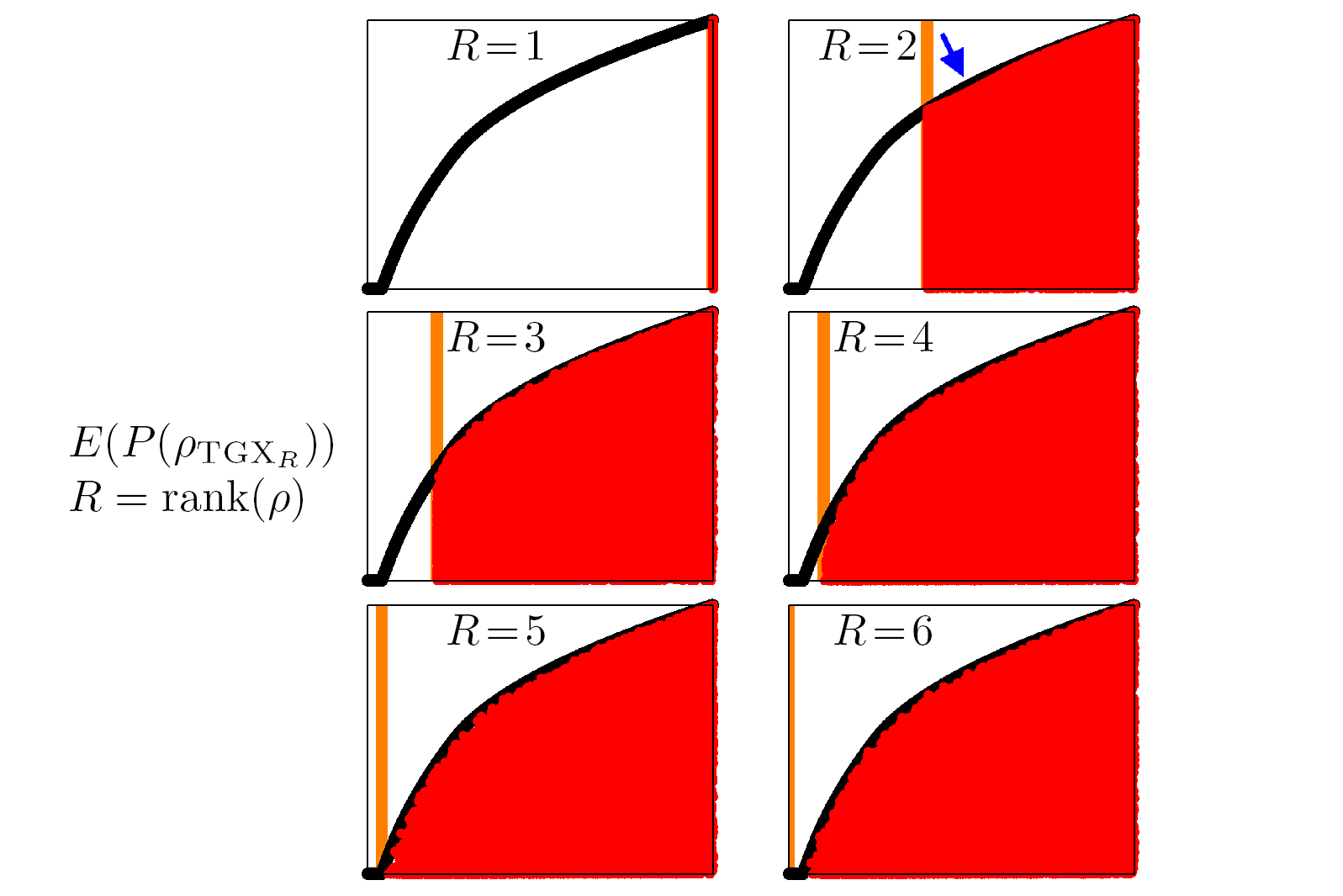}
\caption[]{\label{fig:8}(color online) Entanglement vs.~purity for $50,000$ true-general X states $\rho_{\text{TGX}_R}$ of rank $R$ in $2\times 3$, each with the minimal purity line at $P_{\text{min}}=\frac{1}{R}$.  Again, the numerical values of each axis match those of Fig.~\ref{fig:6}.  Notice that the rank-2 TGX states reach the EP values where the LX states failed in Fig.~\ref{fig:7}.  However, as the blue arrow indicates, there is a slight dip in EP values below the MEMS border.  As Fig.~\ref{fig:9} shows, this behavior happens to \textit{general} states too.}
\end{figure}

\begin{figure}[H]
\centering
\includegraphics[width=0.99\linewidth]{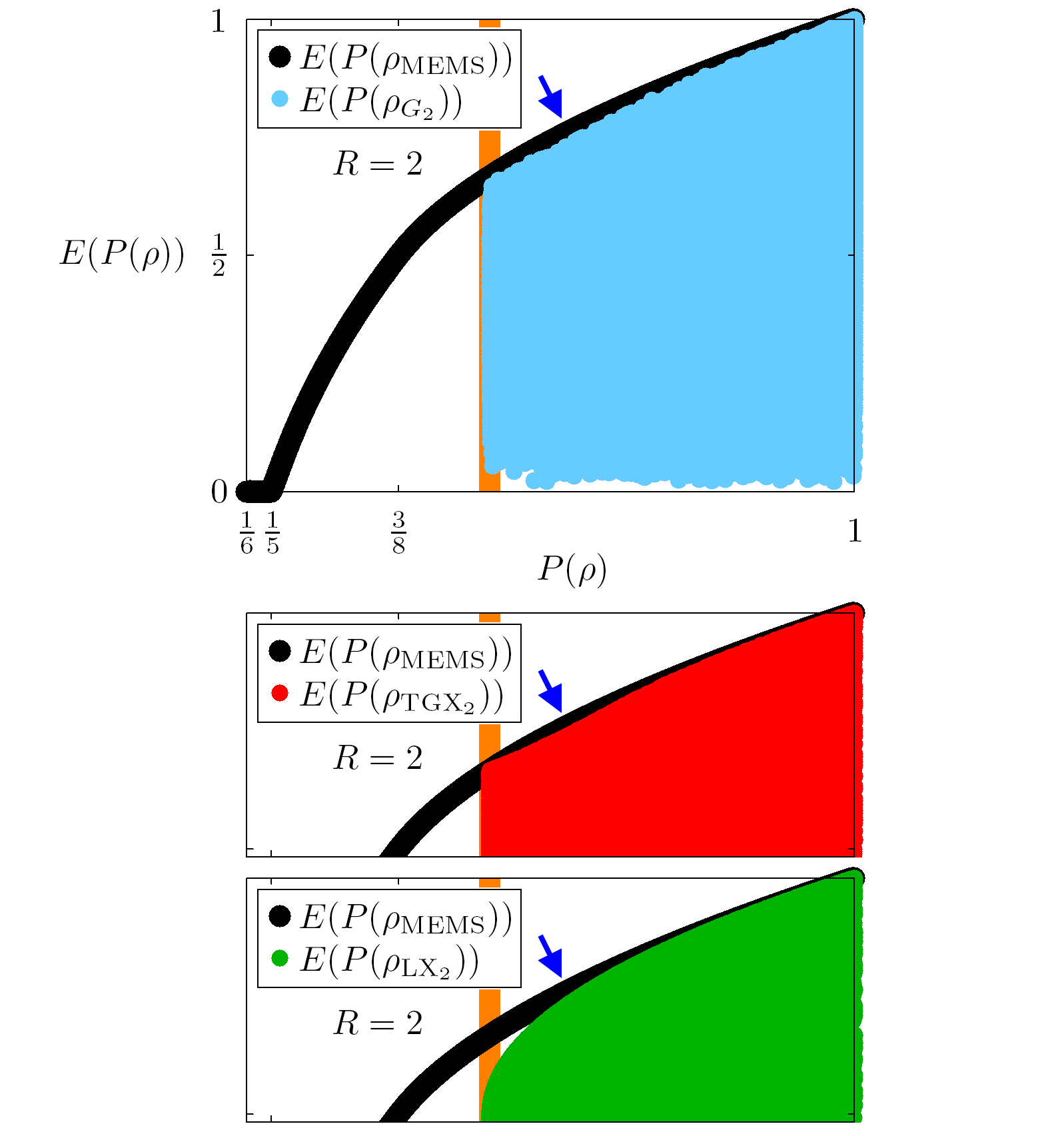}
\caption[]{\label{fig:9}(color online) Entanglement vs.~purity for $2,000,000$ general $2\times 3$ rank-2 states $\rho_{G_2}$ in the top plot.  Although approaching the MEMS curve with random states is difficult, it appears that the general rank-2 states have a slight dip below MEMS, just as the rank-2 TGX states do in Fig.~\ref{fig:8}, shown enlarged in the middle plot here, with the dip clearly visible.  This suggests that the set of rank-2 general states have a concurrence \textit{ceiling} that is slightly \textit{below} the MEMS values.  The rank-2 LX states from Fig.~\ref{fig:7} are also shown enlarged in the bottom plot for a better view.  It was verified that the TGX states completely cover the LX states for rank-2 in all cases.  The bottom two plots are each samples of $100,000$ states, with scale markings matching those of the top plot.}
\end{figure}

Now, we look at TGX states to see if they can do better.  Again, since some combinations perform better than others, only the ones that worked the best are presented here.  Thus, let the rank-specific TGX states be
\begin{equation}
\begin{array}{*{20}l}
   {\rho _{\text{TGX}_1 } } &\!\! { \equiv \rho _{\Phi _1^ +  (\theta _1 )} }  \\
   {\rho _{\text{TGX}_2 } } &\!\! { \equiv p_1 \rho _{\Phi _1^ +  (\theta _1 )}  + p_2 \rho _{\Phi _2^ +  (\theta _2 )} }  \\
   {\rho _{\text{TGX}_3 } } &\!\! { \equiv p_1 \rho _{\Phi _1^ +  (\theta _1 )}  + p_2 \rho _{\Phi _2^ +  (\theta _2 )}  + p_3 \rho _{\Phi _3^ +  (\theta _3 )} }  \\
   {\rho _{\text{TGX}_4 } } &\!\! { \equiv p_1 \rho _{\Phi _1^ +  (\theta _1 )}  + p_2 \rho _{\Phi _2^ +  (\theta _2 )}  + p_3 \rho _{\Phi _3^ +  (\theta _3 )}  + p_4 \rho _{\Psi _1^ +  (\theta _4 )} }  \\
   {\rho _{\text{TGX}_5 } } &\!\! { \equiv\! \left(\! \begin{array}{l}
 p_1 \rho _{\Phi _1^ +  (\theta _1 )}  + p_2 \rho _{\Phi _2^ +  (\theta _2 )}  + p_3 \rho _{\Phi _3^ +  (\theta _3 )}  \\ 
  + p_4 \rho _{\Psi _2^ +  (\theta _4 )}  + p_5 \rho _{\Psi _2^ -  (\theta _5 )}  \\ 
 \end{array}\!\! \right)}  \\
   {\rho _{\text{TGX}_6 } } &\!\! { \equiv\! \left(\! \begin{array}{l}
 p_1 \rho _{\Phi _1^ +  (\theta _1 )}  + p_2 \rho _{\Phi _2^ +  (\theta _2 )}  + p_3 \rho _{\Phi _3^ +  (\theta _3 )}  \\ 
  + p_4 \rho _{\Psi _1^ +  (\theta _4 )}  + p_5 \rho _{\Psi _2^ +  (\theta _5 )}  + p_6 \rho _{\Psi _3^ +  (\theta _6 )}  \\ 
 \end{array}\! \right)\!,}  \\
\end{array}
\label{eq:67}
\end{equation}
where we use the $\theta$ state versions of the states defined in (\ref{eq:49}) and (\ref{eq:50}), and Fig.~\ref{fig:8} shows their EP-plots.
While Fig.~\ref{fig:8} shows that the rank-2 TGX states access the values unreachable to LX states, they have a slight dip below the MEMS line.  To see if \textit{general} states also have this dip, Fig.~\ref{fig:9} plots only rank-2 general states.

If the data in Fig.~\ref{fig:9} is accurate, then the set of rank-2 general states has a concurrence ceiling that is slightly below the MEMS curve, and the ``dip'' seen in both Fig.~\ref{fig:8} and Fig.~\ref{fig:9} is the correct maximum for rank-2 states.

Thus, if rank-2 states truly do have a ceiling slightly less than MEMS, then Fig.~\ref{fig:8} and Fig.~\ref{fig:9} show that \textit{for every EP value accessible to a general state of a given rank, there exists a TGX state of equal rank with that same EP value}.  Therefore, this provides strong evidence that TGX states are the true generalization from the two-qubit case, since every general state can be linked to TGX states by a single EPU matrix.
However, to be fair, in most cases the LX states also have this property, except for the small region of rank 2 states (and perhaps smaller regions for the other ranks that are not visible in these examples).  Yet, since TGX states seem to have no exceptions, they are the more general candidates as universal states, and we may conjecture that this is true regardless of the size or composition of the system, though there may be special cases when LX states can do just as well as TGX states.
\subsection{\label{sec:III.E}Parameterization of General Multipartite States with TGX States}
If TGX states access all entanglement-purity-rank combinations for all $N$-partite systems, as the evidence suggests in Fig.~\ref{fig:2} for $2\times 2$ and in Fig.~\ref{fig:8} for $2\times 3$, then all states should be expressible as a TGX state transformed by an EPU matrix.

Thus, calling the most general real-valued TGX state $\rho_{\text{TGX}_{\text{real}}}$, then the most general multipartite state is
\begin{equation}
\rho = U_{\text{EPU}_{\text{TGX}}} \rho_{\text{TGX}_{\text{real}}} U_{\text{EPU}_{\text{TGX}}}^{\dag},
\label{eq:68}
\end{equation}
where $U_{\text{EPU}_{\text{TGX}}}$ is the most general EPU operation on TGX states, characterized by
\begin{equation}
\begin{array}{*{20}l}
   {U_{\text{EPU}_{\text{TGX}} }\!  \equiv } &\!\!\! \left({\prod\limits_{x = 2}^n {\prod\limits_{y = 1}^{x - 1}\! {U_{(x,y)} } }}\right)\!\!{{\,}^{\dag}}D^{\dag},\;\;\text{s.t.}  \\
   {} &\!\!\! {E(U_{\text{EPU}_{\text{TGX}}} \rho_{\text{TGX}_{\text{real}}} U_{\text{EPU}_{\text{TGX}}}^{\dag})\!=\! E(\rho_{\text{TGX}_{\text{real}}}),\!\!}  \\
\end{array}
\label{eq:69}
\end{equation}
where products are written left to right, $n$ is the total dimension of the multipartite Hilbert space, $E(\rho)$ is any valid multipartite entanglement measure (for \textit{mixed} states), and $U_{(x,y)}$ is an $n$-level single-qubit unitary matrix with one superposition angle and one relative phase angle, where the parenthetical subscripts indicate the subspace upon which $U_{(x,y)}$ acts by referencing its only nonzero off-diagonal element, and $D$ is an $n$-level diagonal unitary matrix.

Thus, in general, $U_{\text{EPU}_{\text{TGX}}}$ intrinsically has $n^2 -1$ DOF to contribute, although since $D$ is next to the TGX state, the DOF can be further reduced by the zeros of $\rho_{\text{TGX}_{\text{real}}}$.

The $2\times 2$ and $2\times 3$ systems are special because each has simple maximally entangled basis (MEB) families consisting of states with only two outcomes each, allowing us to count one superposition angle per pure state in the mixture when we generalize them to $\theta$ states.  However, as we saw in $2\times 2\times 2$, in (\ref{eq:52}) and (\ref{eq:54}), not only are there generally multiple families of simple MEBs with \textit{different numbers of outcomes in superposition}, but also the families with more than two outcomes generalize to ``$\theta$ states'' of \textit{multiple superposition angles per state}.  For that reason, it is possible that the full-rank real-valued TGX states generally have \textit{more} than $2n-1$ DOF, that being the number produced by $n-1$ DOF for the $n$ probabilities, and the minimum of $n$ DOF for the superposition angles of the pure TGX states of the decomposition, \textit{given only two superposition outcomes per state}.

Therefore, treating $2n-1$ as the \textit{minimum} DOF contributed by $\rho_{\text{TGX}_{\text{real}}}$, and ignoring the phase DOF from $D$, then the total DOF of (\ref{eq:68}) would be \textit{at least} $n^2 -n+2n-1=n^2 +n-1$, which is $n$ DOF \textit{more} than the $n^2-1$ DOF needed for a general state, and that is not even counting the DOF contributed by $D$.  Thus, just on the basis of the number of variables that could contribute to the total DOF, it is quite reasonable to hypothesize that there exist $U_{\text{EPU}_{\text{TGX}}}$ that could elevate $\rho_{\text{TGX}_{\text{real}}}$ to a fully general state with $n^2-1$ DOF.

Note that this is just a conjecture, since we do not even have the means to numerically test this for larger systems since no general explicit entanglement measure is known beyond $2\times 3$.  However, the DOF argument does show that there might generally be enough DOF for (\ref{eq:68}) to be valid, and given the successful entanglement-purity-rank plots in Fig.~\ref{fig:8}, it seems at least permissible that (\ref{eq:68}) may be true for all multipartite systems.
\section{\label{sec:IV}Conclusions}
In the first part of this paper, strong numerical evidence was presented in Fig.~\ref{fig:2}, indicating that for two-qubit systems, X states can attain every possible CP value for all possible ranks.  Since we know that unitary operators that preserve entanglement exist, such as the local-unitary matrices $U^{(1)}\otimes U^{(2)}$, this suggests that in principle, all general states may be transformable to X states with a single entanglement-preserving operation.  However, local-unitary matrices are merely \textit{sufficient} for entanglement preservation, since a diagonal and possibly non-separable unitary matrix can also be included, provided that it acts adjacently to the X state.  Therefore, the more general term \textit{entanglement-preserving unitary} (EPU) was introduced to emphasize that the \textit{function} of such matrices as preserving entanglement and purity is more important than their \textit{form}. A means of obtaining this EPU matrix was given explicitly in (\ref{eq:17}), provided that one can find the X state of equal concurrence, purity, and rank, thus proving that, in principle, demonstration that X states can reach all CP values for each rank is necessary and sufficient for proving the claim.  The data collected strongly supports this, and although it is not a proof, it is strong evidence.

To further support the hypothesis that all general two-qubit states are EPU-equivalent to X states, a numerical method of verifying the existence of X-transformations was proposed in Sec.~\ref{sec:II.C.2}.  As shown in Fig.~\ref{fig:3}, the method was found to be successful in transforming general states into X states.  Since this method was successful \textit{every} time for $10,000$ \textit{consecutive} arbitrary input states (only $100$ of which are shown in Fig.~\ref{fig:3} for clarity), and since it relied on the assumption of the existence of an EPU operator to equate the eigenvalues of general states to X states, then this lends strong credibility to the claim of EPU equivalence between general states and X states.

We then briefly examined the fact that this EPU-equivalence, if true, implies that the real ``meat'' of all states, as far as entanglement is concerned, is an X state rotated by a single EPU matrix, the most general form of which was characterized in (\ref{eq:30}).  This is highly analogous to the relationship of all pure states to the set of pure states without global phase.  In that case, the only part we need to care about is the globally-phase-stripped state, and the set of all possible pure states is connected to these by a $1$-level unitary, which is merely a unit-complex phase factor.  Thus, (\ref{eq:29}) is similar to this; the set of general mixed states can be obtained from X states by transforming with a single EPU matrix.  Figure \ref{fig:3} suggests that in fact, the EPU operator $\epsilon_{\rho_X}\epsilon_{\rho_G}^{\dag}$ is all that is needed to reach X form from the general states.  The existence of X transformations is further supported by \cite[]{Lein}, in which it is shown that local-unitary matrices can be used on the Bloch-vector expansion of the state to achieve a Schmidt decomposition, which suggests that it may be possible to transform the new (state-dependent) Schmidt basis to X form with some EPU transformation.  A related result was given in \cite[]{Szal}, showing that a more restricted class of two-qubit states reduced from a larger Fermionic system of qubits can always be local-unitarily transformed to X states.  However, since it is important to observe that not all entanglement-preserving transformations have a product form, we have used the more general EPU terminology where possible in this work.

To conclude the two-qubit discussion, we then observed that for every general mixed state, there is always a pure state of equal concurrence, and pointed out that this should hold for all higher-dimensional systems, as well.  A new entanglement measure was then proposed, by supposing the existence of entanglement-preserving depolarization channels $\Lambda(\rho)$.  Some basic requirements of these channels were discussed and posed as an open problem that may lead to a universal entanglement measure.  Then, to spur research in that direction, we presented a new class of states of parametric entanglement and purity, the H states of (\ref{eq:35}).  Although principally invented because they allow constant $C$ as $P$ varies, both parameters can be varied in any manner desired, with time or otherwise, making these states potentially extremely useful in a wide variety of situations.  Hopefully they will aid in the development of entanglement-preserving depolarization channels.  A major difficulty for that task is the need to be able to change \textit{rank} without changing entanglement.  If such a limitation could be overcome, then the H states of constant $C$, shown in Fig.~\ref{fig:5}, as well as their generalizations in larger systems, could be used to find entanglement in any situation, by relating the entanglement of a general mixed state to that of a pure state.  This is the origin of the idea of \textit{angle of entanglement}, since the superposition angle of the corresponding pure state would correspond to the entanglement value of the arbitrary input state.

The remaining part of this paper was dedicated to showing that the \textit{true} generalization of X form does not have a literal X shape, and that we are justified in claiming this is true because only these true-generalized X states (TGX states) can be EPU-linked with the set of general states by rank.  To define such TGX states operationally, first a method was given to construct them using the requirement that full density matrix elements contributing to off-diagonals of all reductions be identically zero.  Then, an alternate definition was given in terms of exhaustively finding all \textit{simple} complete or overcomplete families of maximally entangled pure states, where ``simple'' was defined as any maximally entangled state fitting the reduction-requirement just mentioned.  This method was shown to agree with the other, with the added bonus that it offers support on a complete or overcomplete set of maximally entangled states.  In contrast, \textit{literal} X states (LX states) do not always contain complete maximally entangled basis sets as in $2\times 3$, or if they do, such as in $2\times 2\times 2$, they do not contain \textit{all} such families, while the TGX states \textit{do} contain all such families.  We may hypothesize that the ability of the TGX-states to contain all families of simple maximally entangled states is what enables them to be EPU-equivalent to all general states.  To be published in 2014, \cite[]{HedT} contains a treatment of TGX states using mulitpartite Bloch vectors.  In that work, an explicit formula is given for generating TGX states in any sized system of any composition.

Finally, to support the claim that TGX states are superior to LX states, the $2\times 3$ system was explored, using a measure similar to negativity based on the PPT test of Peres.  We presented a heuristically-motivated candidate family of $2\times 3$ MEMS in (\ref{eq:64}), and these appear to consistently have the highest possible entanglement in all subsequent plots, lending strong support to the proposition that they are the proper MEMS for $2\times 3$.  To this author's knowledge, these are the first MEMS presented beyond $2\times 2$.  A more methodical means of finding MEMS was presented in \cite[]{Vers}, but that is only for $2\times 2$, and relies specifically on concurrence.  Then, using these MEMS candidates as a reference, we explored rank-specific state families of LX states and TGX states for $2\times 3$, and found that in the case of rank-2 states, the combination of LX states that achieved the highest entanglement could not achieve \textit{all} of the entanglement at purities near the rank-2 minimum near the MEMS curve, as seen in Fig.~\ref{fig:7}.  By contrast, the TGX states appear to successfully achieve all possible entanglement values for all purities and all ranks.  Thus, as far as present evidence suggests, only TGX states can be related to all general states with a single EPU matrix.  Of course, our inability to find a rank-2 LX state that behaves better is not proof that one does not exist.  But for now, it appears that TGX states are the true universal states for $2\times 3$.  We may hypothesize that this is true for all possible systems and compositions.

Furthermore, in the comparison of LX states and TGX states for $2\times 3$, we found that while the TGX states definitely out-perform the LX states in rank-2, they also exhibit a slight ``dip'' below the MEMS line.  This prompted us to take a closer look at the set of \textit{general} rank-2 states in Fig.~\ref{fig:9}, and indeed we found similar behavior there, as well.  Thus, not only does it appear that the TGX states can definitely access all the same EP-values that general states can for each given rank, but the TGX states also allow us to more easily probe the limits of what is possible with general states.

Interestingly, the $2\times 3$ MEMS presented in (\ref{eq:64}) are LX states.  The possibility that $N$-qubit systems posses MEMS of LX form was proposed in \cite[]{Agar}, and although $2\times 3$ is different kind of system, the apparent literal X form of its MEMS suggests that MEMS can always have LX form in all systems, though this is still an open question.

While this paper focused mainly on numerical results and qualitative observations, all are deeply rooted in theoretical analysis and numerical methods.  Although not much here can be taken as solid proof, the results found are in strong support of all hypotheses put forth.  Thus, at the very least, this paper can serve as a guidepost to prompt further research in this area.  It is hoped that the idea of universal states with simpler form than general states may enable algebraic calculations of entanglement in large systems.  For example, \cite[]{EbGM} proved that the genuinely-multipartite concurrence (GM concurrence) \cite[]{MaCh,Love,Pope} of $N$-qubit systems in a literal X form could be computed algebraically.  If the results suggested by this present work are valid, we are guaranteed to be able to transform any general state to TGX form.  Then, as is suggested by (\ref{eq:56}), if the LX states happen to be a subspace of the TGX states, a further entanglement-preserving transformation may also be able to reduce the state to an LX state, thus enabling direct calculation of the GM concurrence for any $N$-qubit input state.

The importance of literal X states (LX states) is already well-known, as many authors have already discovered novel properties arising from systems that initially start in LX form \cite[]{YuE2,AlQa,Wein}, with experimental realization demonstrated in \cite[]{Pete}.  The identification of the TGX states presented here generalizes this class of states to all possible systems, and therefore may represent a universal form for initial states, enabling states to be prepared in the simplest manner possible without sacrificing generality with respect to purity and entanglement.

Furthermore, this paper suggests a new kind of entanglement measure that uses quantum operations to relate arbitrary states to pure states of the same entanglement.  This could open many doors to exciting research, as could the new H states that allow full control over entanglement and purity.

Thus, we have obtained strong evidence that all states are EPU-equivalent to TGX states, and have established several new tools to enable new research in this area.
\begin{acknowledgments}
Special thanks to Ting Yu, Szil{\' a}rd Szalay, and Lin Chen for helpful discussions.  This project was supported by the I{\&}E Fellowship at Stevens Institute of Technology.
\end{acknowledgments}
\begin{appendix}
\section{\label{sec:AppA}}
Here, we will prove that entanglement-preserving unitary (EPU) matrices for X states are generally nonlocal, meaning that they generally do \textit{not} have the form $U^{(1)}\otimes U^{(2)}$.  We will also prove that single-operator X-transformations without entanglement preservation are always possible.  We will then prove that the degrees of freedom for X states transformed by EPU operators is enough to parameterize a general mixed state.
\subsection{\label{sec:AppA.1}Proof that EPU Operators for X States are Generally Nonlocal}
We will start by proving that diagonal unitary operators are generally  nonlocal, and then we will show that such operators are EPU for X states since they do not change the concurrence.  This will prove that EPU operators for X states are generally nonlocal, and give us clues about a general form for such EPU matrices.

First, define a diagonal unitary matrix,
\begin{equation}
D\equiv\text{diag}\{e^{i\eta_{1}},e^{i\eta_{2}},e^{i\eta_{3}},e^{i\eta_{4}}\},
\label{eq:A.1}
\end{equation}
where $\eta_{1},\eta_{2},\eta_{3},\eta_{4}$ are real.  To see that $D$ is generally not factorizable, note that if it were, its form would be
\begin{equation}
D^{(1)}  \otimes D^{(2)}\!  = \text{diag}\{ e^{i(a_1  + b_1 )}\! ,e^{i(a_1  + b_2 )}\! ,e^{i(a_2  + b_1 )}\! ,e^{i(a_2  + b_2 )} \},
\label{eq:A.2}
\end{equation}
where $D^{(1)}  \equiv \text{diag}\{ e^{ia_1 } ,e^{ia_2 } \} $ and $D^{(2)}  \equiv \text{diag}\{ e^{ib_1 } ,e^{ib_2 } \} $ are diagonal unitary matrices in each subsystem, with real parameters $a_1,a_2$ and $b_1,b_2$.  Then, comparing (\ref{eq:A.2}) to (\ref{eq:A.1}) shows that $D$ is separable iff
\begin{equation}
\begin{array}{*{20}c}
   {\eta _1  = a_1  + b_1 ,} &\!\! {\eta _2  = a_1  + b_2 ,} &\!\! {\eta _3  = a_2  + b_1 ,} &\!\! {\eta _4  = a_2  + b_2 ,}  \\
\end{array}
\label{eq:A.3}
\end{equation}
which can be restated as
\begin{equation}
\left( {\begin{array}{*{20}c}
   1 & 0 & 1 & 0  \\
   1 & 0 & 0 & 1  \\
   0 & 1 & 1 & 0  \\
   0 & 1 & 0 & 1  \\
\end{array}} \right)\!\!\left( {\begin{array}{*{20}c}
   {a_1 }  \\
   {a_2 }  \\
   {b_1 }  \\
   {b_2 }  \\
\end{array}} \right)\! =\! \left( {\begin{array}{*{20}c}
   {\eta _1 }  \\
   {\eta _2 }  \\
   {\eta _3 }  \\
   {\eta _4 }  \\
\end{array}} \right)\!.
\label{eq:A.4}
\end{equation}
Then, if $M$ is the transformation matrix in (\ref{eq:A.4}), we see that it is not invertible since
\begin{equation}
\det(M)=0,
\label{eq:A.5}
\end{equation}
and therefore it is \textit{not} always possible to express the local unitary parameters $a_1,a_2,b_1,b_2$ as linear combinations of the general parameters $\eta_{1},\eta_{2},\eta_{3},\eta_{4}$.  The only case when $D$ is factorizable is when $\eta_{1},\eta_{2},\eta_{3},\eta_{4}$ satisfy the conditions in (\ref{eq:A.3}) for real numbers $a_1,a_2,b_1,b_2$.  Thus we have proven that a diagonal unitary matrix $D$ is generally nonfactorizable, and thus nonlocal.

To illustrate this with an example, note that the following equations are necessary conditions for factorizability of diagonal unitary matrices,
\begin{equation}
\begin{array}{*{20}l}
   {\eta _4  - \eta _3 } &\!\! { = \eta _2  - \eta _1, }  \\
   {\eta _4  - \eta _2 } &\!\! { = \eta _3  - \eta _1, }  \\
\end{array}
\label{eq:A.6}
\end{equation}
which can be seen from (\ref{eq:A.3}).  Then, defining an arbitrary diagonal unitary matrix $D$ with angles (in radians),
\begin{equation}
\begin{array}{*{20}c}
   {\eta _1  = 0.95,} &\! {\eta _2  = 0.23 ,} &\! {\eta _3  = 0.61 ,} &\! {\eta _4  = 0.49 ,}  \\
\end{array}
\label{eq:A.7}
\end{equation}
testing these in (\ref{eq:A.6}) shows that they \textit{fail} since 
\begin{equation}
\begin{array}{*{20}r}
   { - 0.12\ne } &\!\! { - 0.72,}  \\
   {0.26\ne } &\!\! { - 0.34,}  \\
\end{array}
\label{eq:A.8}
\end{equation}
Thus, this example shows that diagonal unitary matrices exist that fail to satisfy necessary conditions for factorizability of diagonal unitary matrices, thus verifying the above proof that $D$ is generally nonfactorizable.

Next, applying the generally nonfactorizable $D$ of (\ref{eq:A.1}) to the general X state of (\ref{eq:3}) produces
\begin{equation}
D\rho _X D^{\dag}   =\! \left(\! {\begin{array}{*{20}c}
   {\rho _{1,1} } & 0 & 0 & {\rho _{1,4} e^{ - iA} }  \\
   0 & {\rho _{2,2} } & {\rho _{2,3} e^{ - iB} } & 0  \\
   0 & {\rho _{3,2} e^{iB} } & {\rho _{3,3} } & 0  \\
   {\rho _{4,1} e^{iA} } & 0 & 0 & {\rho _{4,4} }  \\
\end{array}}\! \right)\!,
\label{eq:A.9}
\end{equation}
where the phase factors are functions of parameters of $D$,
\begin{equation}
A \equiv \eta _4  - \eta _{1}\;\; \text{and}\;\;B \equiv \eta _3  - \eta _2 .
\label{eq:A.10}
\end{equation}
First, note that the application of $D$ preserves the X form of $\rho _X $.  Therefore, we can use (\ref{eq:4}) to calculate the concurrence of $D\rho _X D^{\dag}$ algebraically, which shows that
\begin{equation}
\begin{array}{*{20}l}
   {C(D\rho _X D^{\dag}  )} &\!\!  { = 2\max\! \left\{\! \begin{array}{l}
 0,|\rho _{3,2} e^{iB} | - \sqrt {\rho _{4,4} \rho _{1,1} } , \\ 
 |\rho _{4,1} e^{iA} | - \sqrt {\rho _{3,3} \rho _{2,2} }  \\ 
 \end{array}\!\! \right\}}  \\
   {} &\!\!  { = 2\max\! \left\{\! \begin{array}{l}
 0,|\rho _{3,2} | - \sqrt {\rho _{4,4} \rho _{1,1} } , \\ 
 |\rho _{4,1} | - \sqrt {\rho _{3,3} \rho _{2,2} }  \\ 
 \end{array}\!\! \right\}}  \\
   {} &\!\!  { = C(\rho _X )},  \\
\end{array}
\label{eq:A.11}
\end{equation}
thus proving that a generally nonfactorizable diagonal unitary matrix $D$ cannot change the entanglement of an X state.  This means that local-unitarity is merely sufficient for preservation of entanglement of X states.  

Therefore, a more general form (but still not the most general) for EPU operations on X states is given by
\begin{equation}
U_{\text{EPU}_{\text{X}}}  \equiv (U^{(1)}  \otimes U^{(2)} )D,
\label{eq:A.12}
\end{equation}
where $D$ is a diagonal unitary matrix (generally nonfactorizable), and $U^{(m)}$ is a unitary matrix in subsystem $m$, and $D$ \textit{must} be applied adjacently to the X state.  (\ref{eq:A.12}) uses the well-known fact that local-unitary matrices preserve entanglement for all states. Thus, we have proven that a more general form for EPU matrices acting on X states is generally nonfactorizable and thus nonlocal.

However, \textit{other} types of transformations exist that can also preserve entanglement of X states.  For example, consider the nonlocal X-preserving unitary matrix,
\begin{equation}
U_{X} \equiv\! \left(\! {\begin{array}{*{20}c}
   {c_\varepsilon  e^{i\alpha } } & 0 & 0 & {s_\varepsilon  e^{i\beta } }  \\
   0 & {c_\theta  e^{i\phi } } & {s_\theta  e^{i\chi } } & 0  \\
   0 & { - s_\theta  e^{ - i\chi } } & {c_\theta  e^{ - i\phi } } & 0  \\
   { - s_\varepsilon  e^{ - i\beta } } & 0 & 0 & {c_\varepsilon  e^{ - i\alpha } }  \\
\end{array}} \right)\!,
\label{eq:A.13}
\end{equation}
where $c_\theta   \equiv \cos (\theta )$, $s_\theta   \equiv \sin (\theta )$, $\varepsilon,\theta\in[0,\frac{\pi}{2}]$, and $\alpha,\beta,\phi,\chi\in[0,2\pi)$.  In general (\ref{eq:A.13}) does not preserve entanglement, but use of (\ref{eq:4}) can show that there are certain conditions that allow the parameters of $U_{X}$ to preserve the entanglement of any X state upon which it acts.  Thus, a subset of (\ref{eq:A.13}) consists of nonlocal nondiagonal X-preserving unitary matrices that qualify as EPU.

In fact, there are also nonlocal \textit{non}-X-preserving unitary matrices that can qualify as EPU, such as 
\begin{equation}
U_{(3,1)}  \equiv\! \left(\!\! {\begin{array}{*{20}c}
   {c_\theta  e^{i\phi } } & 0 & {s_\theta  e^{i\chi } } & 0  \\
   0 & 1 & 0 & 0  \\
   { - s_\theta  e^{ - i\chi } } & 0 & {c_\theta  e^{ - i\phi } } & 0  \\
   0 & 0 & 0 & 1  \\
\end{array}} \right)\!,
\label{eq:A.14}
\end{equation}
again, subject to the constraint that it preserve the entanglement of the state upon which it acts, except that in these cases, the transformed state will not have X form, and therefore we cannot enjoy the benefit of (\ref{eq:4}).

We may then conclude that for two-qubit systems, \textit{the most general EPU matrix acting on X states is a general-unitary matrix for which its $n^2$ independent unitary parameters have been collectively constrained so that the total operation preserves the entanglement of the input state}. Thus, the most general EPU matrices for two-qubit X states can be expressed as
\begin{equation}
\begin{array}{*{20}l}
   {U_{\text{EPU}_X }  \equiv } &\!\! {(U_{(2,1)} U_{(3,1)} U_{(3,2)} U_{(4,1)} U_{(4,2)} U_{(4,3)})^{\dag}D^{\dag} ;}  \\
   {} &\!\! {\text{s.t.}\;\;E(U_{\text{EPU}_X } \rho _{X} U_{\text{EPU}_X }^{\dag}  ) = E(\rho _{X} ),\;\;\forall x,y,}  \\
\end{array}
\label{eq:A.15}
\end{equation}
where $E(\rho)$ is any valid entanglement measure, and the parenthetical subscript indicates the subspace upon which the two-parameter single-qubit unitary matrix $U_{(x,y)}$ acts by referencing its only nonzero off-diagonal element, such as in (\ref{eq:A.14}), and $D$ is a diagonal unitary matrix ($D$ requires no constraints since it is automatically EPU on X states as proven in (\ref{eq:A.11})).  The two parameters of $U_{(x,y)}$ are its superposition angle and a single relative phase angle, for example, set $\chi=0$ in $U_{(3,1)}$ in (\ref{eq:A.14}).  Note that each successive single-qubit unitary transformation in (\ref{eq:A.15}) does not necessarily preserve the entanglement and is thus not necessarily EPU for its input, but rather the \textit{entire transformation} $U_{\text{EPU}_X }$ is EPU for its input $\rho _{X}$.

While we could certainly obtain the set of constraint equations required in (\ref{eq:A.15}) to obtain a general parameterization of such EPU matrices, it is likely that they would be highly complicated and thus impractical to use.  However, for general purposes, it is enough that Fig.~\ref{fig:3} provides strong evidence that such EPU transformations exist for every possible input state, for which the desired EPU matrix is given explicitly by (\ref{eq:17}), given that we have located the X state of the same $C$, $P$, and $R$ as the general input state, which appears to always be possible based on comparison of Fig.~\ref{fig:2} and Fig.~\ref{fig:1}.

Thus, we have proven that EPU matrices for X states are generally nonlocal and we have concluded that they take a general-unitary form subject to an over-all entanglement preservation constraint, as stated in (\ref{eq:A.15}).
\subsection{\label{sec:AppA.2}Proof that All States Can be Transformed to X States With a Single Unitary Matrix}
Suppose we do not care about entanglement preservation, and instead wish merely to transform an arbitrary general state $\rho_G$ to an X state $\rho_X$ using only a single unitary matrix.  Then, one way to do this is
\begin{equation}
U_X \epsilon _{\rho _G }^{\dag}  \rho _G \epsilon _{\rho _G } U_X^{\dag}   = \rho _{X} ,
\label{eq:A.16}
\end{equation}
where $\epsilon _{\rho _G }$ is the eigenvector matrix of $\rho_G$, and $U_X$ is an X-shaped unitary matrix such as in (\ref{eq:A.13}).  Since  $\epsilon _{\rho _G }$ always exists, and since $\epsilon _{\rho _G }^{\dag}  \rho _G \epsilon _{\rho _G }$ is always diagonal, then $\epsilon _{\rho _G }^{\dag}  \rho _G \epsilon _{\rho _G }$ always has X form trivially.  Since $U_X$ preserves X form but not necessarily diagonal form, then (\ref{eq:A.16}) generally has a nondiagonal X form, and is always possible.  Thus we have proven that if we ignore entanglement, then it is easy to transform any general state to an X state, using the unitary matrix $U_X \epsilon _{\rho _G }^{\dag}$.  This may not be the most general X transformation, but it is enough to prove that such transformations are always possible.  Note that this ability to link general states to X states without regard for entanglement is not as useful as the EPU X transformations that were demonstrated in this paper.  Nevertheless, (\ref{eq:A.16}) provides a simple proof-of-concept for the convertibility of general states to X form.
\subsection{\label{sec:AppA.3}Proof that EPU-Transformed X States Can Have the Full Degrees of Freedom of General States}
First, note that the number of degrees of freedom (DOF) that can be brought to a state by a \textit{transformation} is actually \textit{dependent on the state}.  To see this, the $D$ in (\ref{eq:A.1}) intrinsically has $4$ DOF.  It loses $1$ DOF in the matrix transformation which is immune to global phase, effectively setting $\eta_1 =0$, and leaving $D$ with $3$ DOF.  But then, when $D$ acts on $\rho_X$, only $A$ and $B$ survive, so that $D$ can only bring up to $2$ DOF to $\rho_X$, the phase angles defined in (\ref{eq:A.10}).

Furthermore, since $D$ also preserves X form, the number of DOF it can bring to $\rho_X$ depends on specific phase properties of $\rho_X$.  If $\rho_{X_{\text{real}}}$ is real-valued (having no phase factors), then $D$ can bring up to $2$ DOF to it.  However, if $\rho_X$ is fully-phased (having variable phase angles in both its anti-diagonal elements), then $D$ cannot increase the number of DOF further because the sum of its phase-contributions and the existing phase variables constitutes a new set of two phase variables, leaving the number of DOF unchanged.

Now, we will treat two extreme cases.  Recall from (\ref{eq:14}) that a full-rank \textit{real-valued} X state has $7$ DOF ($3$ DOF for the four probabilities and $4$ DOF for the four different superposition angles).  For a fully-phased X state, recall from (\ref{eq:13}) that a full-rank X state only needed two of its contributing pure states to have phase factors so that fully-phased X states have $9$ DOF.  Summarizing, we have
\begin{equation}
\text{DOF}(\rho_{X_{\text{real}}})=7\;\;\text{and}\;\;\text{DOF}(\rho_{X})=9,
\label{eq:A.17}
\end{equation}
where $\rho_{X_{\text{real}}}$ is a full-rank real-valued X state such as $\rho_{X_{4}}$ from (\ref{eq:14}), and $\rho_{X}$ is a fully-phased X state as in (\ref{eq:13}).

The application of $D$, as described above, causes
\begin{equation}
\text{DOF}(D\rho_{X_{\text{real}}}D^{\dag})=9\;\;\text{and}\;\;\text{DOF}(D\rho_{X}D^{\dag})=9,
\label{eq:A.18}
\end{equation}
where again, note that $D$ cannot increase the DOF of a fully-phased X state, rather it merely changes how the phase DOF are defined.  In fact, as can be seen in (\ref{eq:A.9}), we can think of a fully-phased X state as a real-valued X state transformed by a diagonal unitary matrix $D$, so that in general, $\rho_{X}=D\rho_{X_{\text{real}}}D^{\dag}$.

Next, we consider the fact that local-unitary matrices $U^{(1)}\otimes U^{(2)}$ can only bring up to $6$ DOF to X states, since each single-qubit unitary matrix can bring up to $3$ DOF.  This is provable by actually computing $(U^{(1)}\otimes U^{(2)})\rho_{X}(U^{(1)}\otimes U^{(2)})^{\dag}$, and showing that in general any one of the off-diagonal elements requires all $6$ angles from $U^{(1)}\otimes U^{(2)}$ to define the general element.  Since this is true whether the X state is real-valued or fully-phased, then $U^{(1)}\otimes U^{(2)}$ can bring up to $6$ DOF in either case.  Summarizing, we have
\begin{equation}
\begin{array}{*{20}l}
   {\text{DOF}([\underbrace {(U^{(1)}  \otimes U^{(2)} )}_6\underbrace D_2]\underbrace {\rho _{X_{\text{real}} } }_7[(U^{(1)}  \otimes U^{(2)} )D]^{\dag}  )} &\!\! { = 15}  \\
   {\text{DOF}(\underbrace {(U^{(1)}  \otimes U^{(2)} )}_6\underbrace {\rho _X }_9(U^{(1)}  \otimes U^{(2)} )^{\dag}  )} &\!\! { = 15.}  \\
\end{array}
\label{eq:A.19}
\end{equation}
Finally, since it is well-known that a general two-qubit density matrix has $15$ DOF (which are its Bloch vector components), then (\ref{eq:A.19}) shows that an X state transformed by an EPU matrix can achieve the full number of DOF of a general two-qubit mixed state, thus proving the claim of this section.

However, all we have shown here is that the EPU matrix $(U^{(1)}\otimes U^{(2)})D$ \textit{can} increase the DOF of $\rho _{X_{\text{real}} } $ to $15$, but that does not mean those DOF are enough for the transformed state to reach \textit{all} possible general states.

The most general EPU $U_{\text{EPU}_{X}}$ was already stated in (\ref{eq:A.15}), and is a constrained general unitary matrix.  Since such an EPU matrix can be written with its $D$ adjacent to the X state, we can consider only a fully-phased X-state of $9$ DOF as input to the six single-qubit unitaries $U_{(x,y)}$, which intrinsically have $n^2 -n =12$ DOF, as defined.  This means that the total entanglement-preservation constraint and the effects of the matrix multiplication should reduce the DOF contributed by $U_{\text{EPU}_{X}}$ down to just $6$ DOF, so that
\begin{equation}
\text{DOF}(\underbrace {U_{\text{EPU}_X } }_{6}\underbrace {\rho _{X } }_{9}U_{\text{EPU}_X }^{\dag}  ) = 15,
\label{eq:A.20}
\end{equation}
where again, if $U_{\text{EPU}_{X}}$ is given by its most general form in (\ref{eq:A.15}), the $D$ cannot increase the DOF of the fully-phased X state further since that would only change the definitions of the phase variables.

One reason we are justified in saying that the single-qubit unitary factors of $U_{\text{EPU}_{X}}$ from (\ref{eq:A.15}) can only collectively contribute up to $6$ DOF to $U_{\text{EPU}_X }\rho _{X }U_{\text{EPU}_X }^{\dag}$ is that the local-unitary product $U^{(1)}\otimes U^{(2)}$ is a \textit{subset} of $U_{\text{EPU}_{X}}$.  Thus, we are guaranteed that $U_{\text{EPU}_{X}}$ can \textit{at least} bring $6$ DOF to $\rho_X$ due to the universal EPU property of local unitaries.  Then, since a general state can have \textit{no more} than $15$ DOF total, we know that $U_{\text{EPU}_{X}}$ can bring \textit{no more than} $6$ DOF to $\rho_X$, and therefore the most general $U_{\text{EPU}_{X}}$ can bring up to $6$ DOF to $\rho_X$.  

However, the important key point here is actually not the $6$ DOF brought to $\rho_X$ by $U_{\text{EPU}_{X}}$, but rather it is the \textit{way} in which it is done.  The more general $U_{\text{EPU}_{X}}$ has more flexibility in how it contributes the $6$ DOF than does the local unitary $U^{(1)}\otimes U^{(2)}$.  Therefore, since $U_{\text{EPU}_{X}}$ is the most general way in which we can contribute the necessary $6$ DOF, and since the numerical evidence in the $10,000$ consecutive successes of the test in Fig.~\ref{fig:3} all show nontrivial general unitary single-qubit factorizations, we can confidently say that (\ref{eq:A.15}) describes the most general form that an EPU can take, making it more general than the simple example in (\ref{eq:A.12}).

Thus, (\ref{eq:A.15}) is the most general EPU operation on a two-qubit X state, and the evidence strongly suggests that the $15$ DOF accessible through such a transformation are always enough for the set of X states to be linked to the set of general states by a single EPU operation.
\end{appendix}
%
\end{document}